\documentclass[twocolumn,showpacs,preprintnumbers,superscriptaddress,amsmath,amssymb]{revtex4-1}
\usepackage{amssymb}		%math formula
\usepackage{amsmath}
\usepackage{graphicx}
\usepackage[normalem]{ulem}
\usepackage{multirow}
\usepackage{appendix}
\usepackage{CJK}
\usepackage[usenames]{color}
\usepackage{bm}
\usepackage[colorlinks,linkcolor=blue,urlcolor=blue,citecolor=blue]{hyperref}
\usepackage{booktabs} 	% table!
\usepackage{leftidx}

\usepackage{soul,xcolor}
\setstcolor{red}

\usepackage{tikz,xcolor,hyperref}

\definecolor{lime}{HTML}{A6CE39}
\DeclareRobustCommand{\orcidicon}{
	\begin{tikzpicture}
	\draw[lime, fill=lime] (0,0) 
	circle [radius=0.16] 
	node[white] {{\fontfamily{qag}\selectfont \tiny ID}};
	\draw[white, fill=white] (-0.0625,0.095) 
	circle [radius=0.007];
	\end{tikzpicture}
	\hspace{-2mm}
}
\foreach \x in {A, ..., Z}{%
	\expandafter\xdef\csname orcid\x\endcsname{\noexpand\href{https://orcid.org/\csname orcidauthor\x\endcsname}{\noexpand\orcidicon}}
}

\makeatletter

\newcommand{\Rmnum}[1]{\expandafter\@slowromancap\romannumeral #1@}
\makeatother
\begin{document}
\begin{CJK*} {UTF8} {gbsn}
%\begin{CJK*} {GB} {gbsn}

\title{Azimuthal-sensitive three-dimensional HBT radius in Au-Au collisions at $E_{beam} = 1.23$$A$ GeV by the IQMD model}

\author{Ling-Meng Fang(房灵猛)}
\affiliation{Shanghai Institute of Applied Physics, Chinese Academy of Sciences, Shanghai 201800, China}
\affiliation{University of Chinese Academy of Sciences, Beijing 100049, China}
% adviser name
\author{Yu-Gang Ma(马余刚)\orcidC{}}\thanks{Email:  mayugang@fudan.edu.cn}
\author{Song Zhang(张松)\orcidB{}}\thanks{Email: song\_zhang@fudan.edu.cn}
\affiliation{Key Laboratory of Nuclear Physics and Ion-beam Application (MOE), Institute of Modern Physics, Fudan University, Shanghai 200433, China}
\affiliation{Shanghai Research Center for Theoretical Nuclear Physics， NSFC and Fudan University, Shanghai 200438, China}

% Part 0.0: abstract
\date{\today}
\begin{abstract}
We used an Isospin dependent Quantum Molecular Dynamics (IQMD) model to simulate Au + Au collisions at beam energy $E_{beam}$ = 1.23$A$ GeV, which corresponds to center of mass energy $\sqrt{s_{NN}} = 2.4$ GeV. Firstly, we obtained reasonable rapidity and transverse mass spectra of $\pi^-$ and $\pi^+$ as well as ``apparent" temperature parameters in comparison with the HADES data. Then by calculating three-dimensional Hanbury Brown and Twiss (HBT) radius of same charged pion-pairs, we obtained the square difference of outward radius and sideward radius, i.e. $R_{out}^{2}-R_{side}^{2}$, as well as the freeze-out volume ($V_{fo}$), which were basically consistent with the trend of HADES experimental results. The azimuthal dependence of the HBT radii was also calculated by a matrix method, and corrected by the rotation matrix. In addition, the eccentricities of the $xy$- and $zy$- planes of pion-pair emissions were also extracted for different pair transverse momentum and collision centrality,
which show that the $xy$-eccentricity increases with pair transverse momentum as well as centrality but not for $zy$-eccentricity, indicating a more asymmetric transverse emission of particle-pairs with higher  transverse momentum in off-central collisions.

 	\end{abstract}
\maketitle

	% Part 1.0 introduction
	\section{Introduction}
	% basic intro

       Two-particle hadronic intensity interferometry has been widely used to study the phase space evolution of the source in heavy-ion reactions~\cite{ref1,ref2,ref3}. The technique, first proposed by Hanbury Brown and Twiss (HBT), was based on the quantum statistical interference of identical particles~\cite{ref4,ref5} and   was  used to measure angular radii of stars in the beginning and later became known as HBT interference. Goldhaber {\it et al.} applied such an intensity interference method to hadronic systems~\cite{ref6,ref7}. However, considering that the shape and size of the particle source is changing due to the influence of space-momentum correlation during  collisions~\cite{ref8}, intensity interferometry usually does not yield an exact particle source size, but rather an approximately valid ``uniformity length". Intensity interferometry measures the source regions where the momenta of particle pairs are relatively close to each other, so they are correlated due to quantum statistics or two-body interactions~\cite{ref9,ref10}. In the  region of collision energy at GeV per nucleon, the measured particles usually come from  different processes, and intensity interference method can provide additional information to understanding the production mechanism  and dynamics of the particle emission source~\cite{ref11,ref12,ref13,LiLY}.
     
      The intensity interference of pions in heavy-ion collisions was first observed in the Berkeley Bevalac experiment~\cite{ref14} where  an oblate shape of the pion source and a correlation of the source size with the system size were also found~\cite{ref15}. However, people realized that the HBT radius is not only directly corresponding to the spatial size of source, but also related to the so-called ``region of homogeneity". On the other hand, HBT correlation can be used to measure the spatial and temporal information of particle emission source in heavy-ion collisions, which can be used to extract the dynamic information of emission sources. Its related experimental measurement and theoretical analysis have therefore aroused great interests~\cite{ref_nature_2007,ref_nature_star,Neha,ref_nature_alice,ref16,ref17,ref_plum,ref18,ref18B,ref_wangtt,ref_zhou,XiBS}.
   
      Since the 1990s, many parametric models have been developed for a more visualized understanding and physical information extraction of the emission sources. 
      %~\cite{ref19,ref20,ref21}
      On the basis of certain assumptions, some parameterized models can obtain simple and intuitive physical analytical formulas for the HBT radius. This provides a great help in studying the relation between the particle emission and final-state particles \cite{ref_nature_star,Neha,ref_nature_alice}.
      
      In non-central heavy-ion collisions, the random distribution of symmetrical axis of each nucleus-nucleus collision results in smearing or weaken azimuthal angular dependence of the HBT radius. Some works proposed the so-called matrix method~\cite{ref59,HADES} to correct this effect through simulating the coordinates of the initial nucleons. In this work, we also consider if this correction can be obtained by momentum distributions at final state, which will be more convenient in experiment with a 4$\pi$ detector.
      
      This paper is organized as follows. First, a concise introduction to the IQMD model as well as the HBT calculation method is  given in Sec.~\ref{sec:model}. Next, the pion yields and three-dimensional pion HBT radius results as well as the corresponding results with experimental data are presented in Sec.~\ref{sec:analysis}. The azimuthal-angle dependent HBT radii before and after the corrections as well as the eccentricities of $xy-$ and $zy-$ planes  
      are also given in this section. Finally, a brief summary is presented in Sec.~\ref{sec:summary}.

	\section{Models and methods}
	\label{sec:model}
   
       In experiments, particles are measured in heavy-ion reactions at final stage and it is impossible to directly trace the physical evolution in reactions, so  various  models have been  developed to simulate the reaction process. %~\cite{ref22,ref23}. 
       The commonly-used heavy-ion reaction models can be basically divided into statistical models and transport models. %~\cite{ref24}. 
       In this study, an Isospin dependent Quantum Molecular Dynamics (IQMD) model, a kind of transport models,  is employed to study the reaction system from initial state to final stage in low and medium energy heavy-ion collisions. The HBT radii are calculated from the phase-space data of pions at the freeze-out stage simulated by the IQMD model~\cite{PRATT1994103}.  In the following, the IQMD model and HBT calculation methods will be introduced, separately.

	\subsection{The IQMD model}
	
      Quantum Molecular Dynamics  (QMD) model is a molecular dynamics model, from which we can get the information on both the collision dynamics and the phase space information~\cite{ref_aichelin,ref_ma,ref_feng,ref_feng2,ref_guo}. The IQMD model is based on the traditional  QMD model, by including the isospin degree of freedom of nucleons~\cite{ref38}.

      {In the IQMD model, the normalized wave function of each nucleon is expressed in the form of a Gaussian wave packet, 
      \begin{equation}
      \phi_{i}(\vec{r}) = \frac{1}{(2\pi L)^{3/4}}\exp(\frac{-(\vec{r}-\vec{r_{i}}(t))^{2}}{4L})\exp(\frac{i\vec{r}\cdot \vec{p_{i}}(t)}{\hbar}),
      \end{equation}
      here $\vec{r_{i}}(t)$ and $\vec{p_{i}}(t)$ are time-dependent variables describing the center of the wave packet in coordinate space and momentum space, respectively. Given the direction of $\vec{r_{i}}$ and $\vec{p_{i}}$, $\phi_{i}(\vec{r})$ is a six-dimensional function. The parameter $L$ is the width of the wave packet, which is related to the size of the reaction system and usually fixed in the simulations. Here the width $L$ is fixed as $2.16fm^{2}$ for Au + Au reactions ~\cite{PhysRevC.96.064604,ref_yu}.}
      
     All the nucleons interact through an effective mean field and two-body scatterings. The interaction potential can be expressed as
      \begin{equation}
      U = U_{Sky} + U_{Coul} + U_{Yuk} + U_{Sym} + U_{MDI} + U_{Pauli},
      \end{equation}
      where $U_{Sky}$, $U_{Coul}$, $U_{Yuk}$, $U_{Sym}$, $U_{MDI}$, and $U_{Pauli}$ represent the density-dependent Skyrme potential, Coulomb potential, Yukawa potential, isospin asymmetric potential, the momentum-dependent interaction potential and Pauli potential, respectively. The nucleon-nucleon collision cross section in the medium ($\sigma_{NN}^{med}$) can be expressed as taken in Refs.~\cite{ref_westfall,ref40,ref_zhang}
      \begin{equation}
      \sigma_{NN}^{med} = (1-\eta \frac{\rho}{\rho_{0}})\sigma_{NN}^{free},
      \end{equation}
      where $\rho_{0}$ is the density of normal nuclear matter, $\rho$ is the local density, $\eta$ is the in-medium correction factor, which is chosen as 0.2 in this paper to better reproduce the flow data~\cite{ref_wangtt}, and $\sigma_{NN}^{free}$ is the free nucleon-nucleon cross section.

      \subsection{The HBT radius}
    
      In heavy-ion collisions, the momentum correlation between two identical particles is widely used to extract the spatial and temporal dynamic information of particle emission source~\cite{ref42,ref43,ref44}. This intensity interference  can be expressed as~\cite{ref45,ref46,ref47,ref48}

       \begin{equation}
      C(\bold{q},\bold{K}) = 1  + \frac{|\int d^{4}x S(x,\bold{K})e^{iq\cdot x}|^{2}}{|\int d^{4}x S(x,\bold{K})|^{2}},
      \end{equation}
      where  $q = p_1 - p_2  = (\bold{p_{1}}-\bold{p_{2}}, E_1-E_2) = (\bold{q}, E_1-E_2)$ and  $K  = \frac{1}{2}(p_{1} + p_{2}) = (\bold{K}, K_0)$ represent the relative momentum and mean pair momentum with $\bold{p_1}$ and $\bold{p_2}$ being the momenta of the two particles. The correlation function  $C(\bold{q},\bold{K})$  relates to the phase-space density of the source function $S(\bold{x},\bold{K})$. In  experimental measurements,  $C(\bold{q},\bold{K})$ is usually parametrized in terms of the intercept $\lambda(\bold{K})$ and the HBT radii $R_{ij}^{2}(\bold{K})$ by
      \begin{equation}
        C(\bold{q},\bold{K})  = 1 + \lambda(\bold{K})\times \exp[-\sum_{i,j=o,s,l}q_{i}q_{j}R_{ij}^{2}(\bold{K})].
      \label{eq:HBTnihe}
      \end{equation}
   
      In the Cartesian $OSL$ system, the relative momentum is decomposed into components parallel to the beam direction ($L$ = longitudinal) and to the pair transverse momentum $\bold{p_{t,12}}  = \bold{p_{t,1}} + \bold{p_{t,2}}$ direction ($O$ = outward), as well as perpendicular to the first two directions ($S$ = sideward)~\cite{ref49,ref50,ref51,ref52,ref53}.
  
      In the last decade, much attention has been paid to the differential measurement of HBT radius from two-particle correlation, and the $\Phi$ (azimuth angle) - dependence has revealed additional information on the space-time structure of the source and given rise to new insights on the underlying nature of the flows~\cite{ref54,ref55}.
 
      The azimuth-sensitive HBT analysis includes six parameters $R_{ij}^{2}$, all of which are functions of three components of the pair transverse momentum $K_{\bot}$, rapidity $Y$ and azimuth angle $\Phi$. They provide the space-time information of the source according to following equations~\cite{ref8,ref56,ref57}:
      \begin{eqnarray}
	 \begin{aligned}
      &R_{s}^{2}(K_{\perp},\Phi,Y) = S_{11}\sin^{2}\Phi+S_{22}\cos^{2}\Phi-S_{12}\sin2\Phi\\
      &R_{o}^{2}(K_{\perp},\Phi,Y) = S_{11}\cos^{2}\Phi+S_{22}\sin^{2}\Phi-S_{12}\sin2\Phi\\&\qquad\qquad\qquad-2\beta_{\perp}S_{01}\cos\Phi-2\beta_{\perp}S_{02}\sin\Phi+\beta_{\perp}^{2}S_{00}\\
      &R_{os}^{2}(K_{\perp},\Phi,Y) = S_{12}\cos2\Phi+\frac{1}{2}(S_{22}-S_{11})\sin2\Phi\\&\qquad\qquad\qquad+\beta_{\perp}S_{01}\sin\Phi-\beta_{\perp}S_{02}\cos\Phi\\
      &R_{l}^{2}(K_{\perp},\Phi,Y) = S_{33}-2\beta_{l}S_{03}+\beta_{l}^{2}S_{00}\\
      &R_{ol}^{2}(K_{\perp},\Phi,Y) = (S_{13}-\beta_{l}S_{01})\cos\Phi-\beta_{\perp}S_{03}\\&\qquad\qquad\qquad+(S_{23}-\beta_{l}S_{02})\sin\Phi+\beta_{l}\beta_{\perp}S_{00}\\
      &R_{sl}^{2}(K_{\perp},\Phi,Y) = (S_{23}-\beta_{l}S_{02})\cos\Phi-(S_{13}-\beta_{l}S_{01})\sin\Phi,
      \label{eq:R2_KT_Phi_Y}
      \end{aligned}
      \end{eqnarray}
      where $\beta_{\perp} = K_{\perp}/K_0$ represents the pair velocity in the transverse direction, and $\beta_{l} = K_{l}/K_0$ represents the pair velocity in the longitudinal directions. $m_{t} = \sqrt{K^{2}_{\perp} + m^{2}_{\pi}}$ is the transverse mass. $S_{\mu\nu}$ denotes the spatial correlation tensor
      \begin{equation}
      S_{\mu\nu} = \langle \tilde{x}_{\mu}\tilde{x}_{\nu} \rangle, \tilde{x}_{\mu} = x_{\mu}-\bar{x}_{\mu}, ~~(\mu,\nu = 0, 1, 2, 3),
      \label{eq:smx}
      \end{equation}
      which measures the Gaussian width in space-time of the emission function $S(\bold{x},\bold{K})$ around the point of highest emissivity $\bar{x}_{\mu} = \langle\tilde{x}_{\mu}\rangle$. $S_{\mu\nu}$ is defined in terms of Cartesian coordinates in an impact parameter fixed system, in which $x_{0} = t$ is the time parameter, $x_{1} = x$ is parallel to the impact parameter b, $x_{3} = z$ is the beam direction, and $x_{2} = y$ is perpendicular to the reaction plane constructed from the $x$ and $z$ directions. The brackets indicate the average over the entire emission source.

      As stated in Refs.~\cite{ref58,ref59}, all elements, except $S_{13}$ and diagonal elements, oscillate symmetrically near zero in collisions with the same mass. This means
      \begin{equation}
      S_{01} = S_{02} = S_{03} = S_{12} = S_{23} = 0.
      \label{eq:Sjuzhen}
      \end{equation}

      As a result, using the fact that the average $\beta_{l}$ is zero near the midrapidity (although average $\beta_{l}^{2} \neq 0$) allows us to express the HBT radii in terms of five non-vanishing components as in Refs.~\cite{ref60,ref61,ref62},
       \begin{eqnarray}
	 \begin{aligned}
      &R_{s}^{2} = \frac{1}{2}(S_{11}+S_{22})+\frac{1}{2}(S_{22}-S_{11})\cos2\Phi\\
      &R_{o}^{2} = \frac{1}{2}(S_{11}+S_{22})-\frac{1}{2}(S_{22}-S_{11})\cos2\Phi+\beta_{\perp}^{2}S_{00}\\
      &R_{os}^{2} = \frac{1}{2}(S_{11}-S_{22})\sin2\Phi\hfill\\
      &R_{l}^{2} = S_{33}+\beta_{l}^{2}S_{00}\\
      &R_{ol}^{2} = S_{13}\cos\Phi\\
      &R_{sl}^{2} = -S_{13}\sin\Phi .
      \label{eq:R2_2Phi}
       \end{aligned}
      \end{eqnarray}
    
      While the amplitude of oscillations of $R_{o}^{2}, R_{s}^{2}$, and $R_{os}^{2}$ are given by the difference between the transverse source sizes in and perpendicular to the reaction plane, that of the oscillations of $R_{sl}^{2}$ and $R_{ol}^{2}$ is given by $S_{13} = \langle\tilde{x}\tilde{z}\rangle$. By parameterizing the emission source by an ellipsoid, a nonzero $S_{13}$ corresponds to a tilt of its longitudinal major axis away from the beam direction in the reaction plane. It can be characterized by a tilt angle,
      \begin{equation}
      \theta_{s} = \frac{1}{2}\tan^{-1}(\frac{2S_{13}}{S_{33}-S_{11}}).
      \label{eq:mx}
      \end{equation}

      Rotating the spatial correlation tensor $S_{\mu\nu}$ by $-\theta_{s}$ yields a purely diagonal tensor $S^{diag} = G_{y}^{\dag}(\theta_{s})\cdot S\cdot G_{y}(\theta_{s})$ with the squared lengths of the three major axes given by its eigenvalues. Using {Eq.~(\ref{eq:R2_2Phi})}, we can then get the HBT radius after the rotation correction.

      \section{Analysis and discussion}
      \label{sec:analysis}
      \par
      In this paper, we use an IQMD model to simulate Au + Au collisions at the beam energy $E_{beam}$ = 1.23$A$ GeV, which corresponds to a center of mass energy $\sqrt{s_{NN}}$ = 2.4 GeV. The total number of events included in the simulation is 1600000.  The centrality is characterized as $c= (\pi b^{2})/(\pi b_{max}^{2})$, where $b$ is the impact parameter, and $b_{max} = 1.15(A_{P}^{1/3} + A_{T}^{1/3})$ is the sum of effective shape radius of projectile and target. The compared HADES experimental results are obtained from Refs.~\cite{HADES,ref56}.

      \subsection{The pion yield}
      \par
       Pions carry interesting information about the high density behavior of symmetry energy and the $\pi^-/\pi^+$  ratio in heavy-ion collisions depends strongly on isospin asymmetry of the reaction system \cite{BALi}. In heavy-ion collisions at beam energies  around about 1 GeV/nucleon and below, most pions are produced through the decay of $\Delta (1232)$ resonances, see, e.g., Refs. \cite{BALi2}, since the direct pion production is very small in the intermediate-energy range and can be neglected.
      
\begin{figure}[htb]
      \includegraphics[angle=0,scale=0.8]{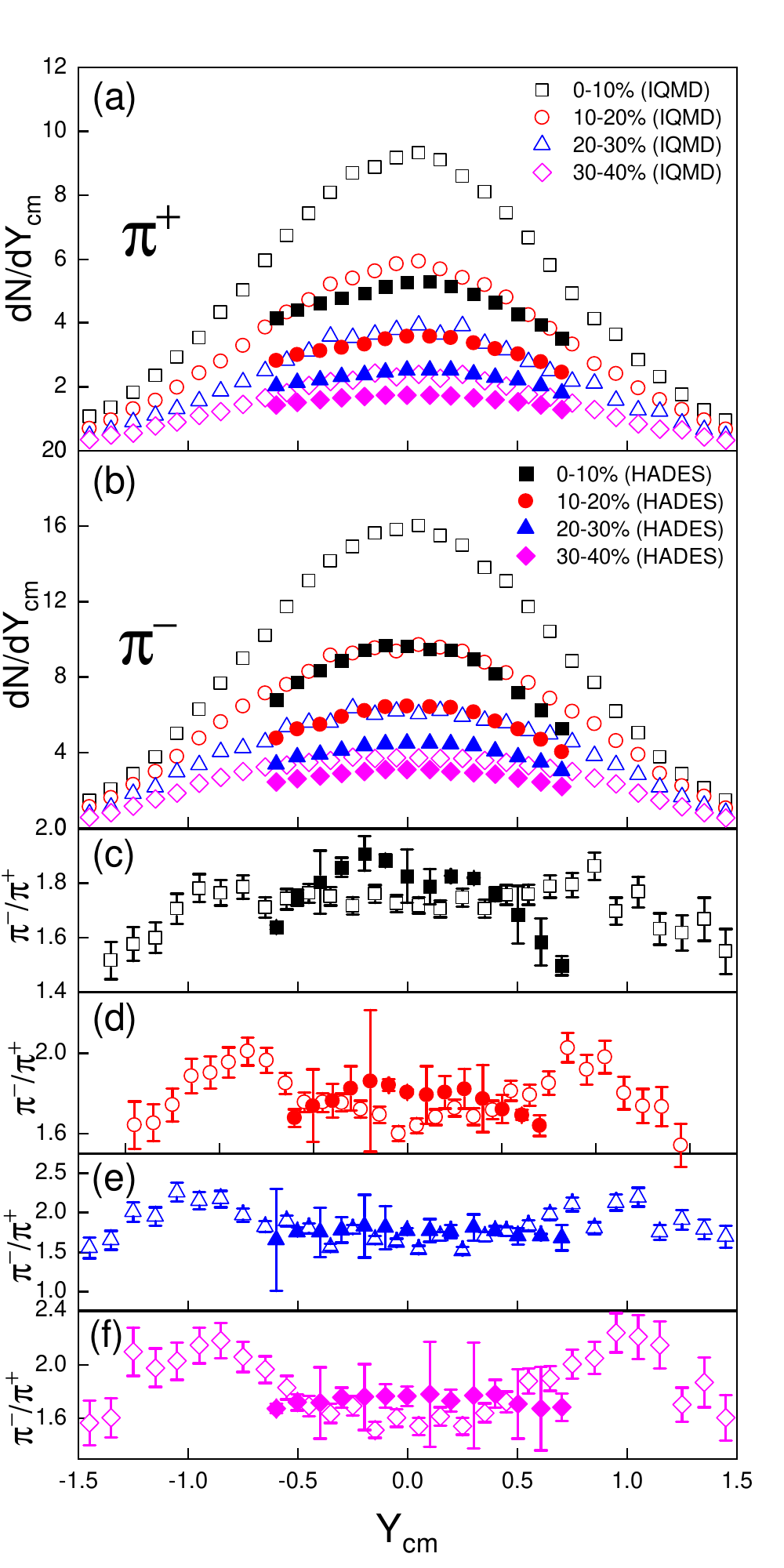}
      \caption{Rapidity distribution of positive (a) and negative (b) charged pions in Au+Au collisions at $E_{beam}$ = 1.23$A$ GeV for four centrality classes with 10\% bin width. The ratios of $\pi ^{-}$ and $\pi ^{+}$ in four centralities are displayed in (c, d, e and f). The solid and hollow points represent the HADES experimental data and the IQMD simulation results, respectively. }
      \label{fig1}
      \end{figure}

\begin{figure}[htb]
      \includegraphics[angle=0,scale=0.42]{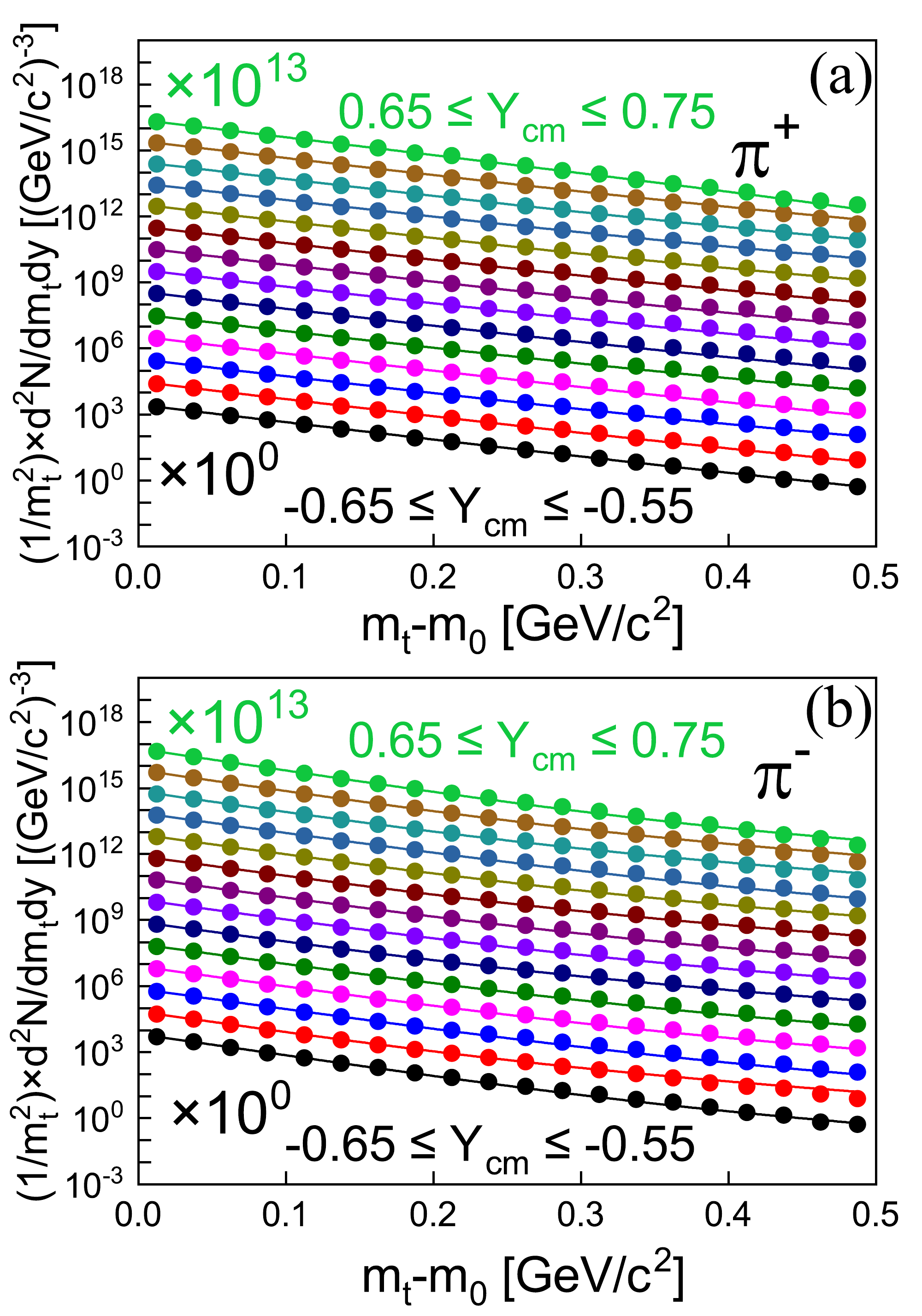}
      \caption{Reduced transverse-mass distributions of positive (a) and negative (b) charged pions in rapidity bins of $\Delta Y_{cm} = 0.1$ width between -0.65 and 0.75 for the 0-10\% most central events from the IQMD model. The spectra in the most backward rapidity region  (i.e., -0.65 $\leq Y_{cm} \leq$  -0.55) are shown un-scalely while the spectra for other rapidity slices are scaled up by successive factors of 10. The solid curves correspond to fits using the two-slope Boltzmann function given by Eq.~(\ref{eq:boltzmann fit}).}
      \label{fig2}
      \end{figure}

      Fig.~\ref{fig1} shows rapidity distributions of positive and negative charged pions for the centrality classes of 0-10, 10-20, 20-30, and 30-40\%. Fig.~\ref{fig2} shows the reduced transverse mass distributions of positive and negative charged pions in different rapidity ranges from $-0.65\leqslant Y_{cm}\leqslant -0.55$ to $0.65\leqslant Y_{cm}\leqslant 0.75$. Also shown by solid lines are fitted results using the following function:
      \begin{equation}
    \frac{1}{m_{t}^{2}}\frac{d^{2}N}{dm_{t}dY_{cm}} = C_{1} exp(-\frac{m_{t}c^{2}}{T_{1}}) + C_{2}exp(-\frac{m_{t}c^{2}}{T_{2}}),
      \label{eq:boltzmann fit}
      \end{equation}
       where the parameters $T_{1}$ and $T_{2}$ account for different slopes at low and high reduced transverse masses, respectively, which we call them ``apparent" temperature parameters, and $C_1$ and $C_2$ relate to the corresponding yields. Here we  chose a superposition of two Boltzmann distributions because the reduced transverse-mass spectrum of pions deviate from a single exponential spectrum. The detailed fitting processes can be found  in Ref.~\cite{HADES}.
      \par
      It is seen from Fig.~\ref{fig1} and Fig.~\ref{fig2} that the yields of $\pi^{+}$ and $\pi^{-}$ gradually decrease
      with less central collision and  larger transverse mass, respectively. We notice that the yield of pions from the IQMD model is higher than experimental data from Fig.~~\ref{fig1}, and the ratio of experimental over model values is $\frac{dn/dY_{HADES}}{dn/dY_{IQMD}}\arrowvert _{\pi^{+}} = 0.54, \frac{dn/dY_{HADES}}{dn/dY_{IQMD}}\arrowvert _{\pi^{-}} = 0.58$. This behavior is basically consistent with the values of $0.523|_{\pi^{+}}$ and $0.566|_{\pi^{-}}$ in previous IQMD results or other models as given in Ref.~\cite{HADES}. As discussed in Ref.~\cite{HADES}, the origin of the displayed overshoot of $\pi$ production remains unclear. Recent calculations on the width of nuclear dipole resonance have demonstrated that the nuclear mean free path is strongly corrected in nuclear medium \cite{WangRui,PhysRevC.96.064604}. Also, it is shown in Ref.~\cite{2107.13384} that the pion yield in this reaction can be described with a reduced nucleon-nucleon  inelastic cross section in medium. However, introducing in-medium corrections of the nucleon-nucleon cross section has little impact on the pion spectrum and its nuclear modification factor \cite{Lv1}. The reason might be that the cross section for pion interaction has no significant change while the in-medium correction for nucleon-nucleon cross section  changes. Therefore  the $\pi$-excess  remains unclear and  deserves further exploration for pion dynamics in near future, eg. the pion (re)scattering and (re)absorption processes, at least in the IQMD framework.

      As mentioned before, the $\pi^-$/$\pi^+$ ratios strongly depend on the isospin of the system, such that $\pi^-$ are produced more abundantly than $\pi^+$ for $^{197}$Au systems. In the $\Delta$ resonance model for pion production from first-chance independent nucleon-nucleon collisions \cite{Stock}, the primordial  $\pi^-$/$\pi^+$ ratio is $(N_{part}/Z_{part})^2$ where $N_{part}$ and $Z_{part}$ are neutron and proton numbers in the participant region of the reaction.  However, because pion reabsorption and rescattering ($\pi + N \leftrightarrow \Delta$ and $N + \Delta \leftrightarrow N + N$) reduce the sensitivity of the $\pi^-$/$\pi^+$ ratio to $(N/Z)_{system}$ as seen in the following.
      In Fig.~\ref{fig1}, a comparison of the  $\pi^-$/$\pi^+$ ratio between our calculation and the HADES mid-rapidity data for four centrality classes is also plotted. Unlike the single rapidity spectra of $\pi^+$ and $\pi^-$, the $\pi^-$/$\pi^+$ ratio cancels out the over-predicted behavior in single spectra, and they can essentially match up with the data within the systematic uncertainty.  For the larger rapidity region (eg. around  ${Y_{cm} = \pm 1.5}$) or off-central collisions (eg. 30-40\% centrality), the values of our calculated $\pi^-/\pi^+$  approach to the initial value of the $(N/Z) \simeq 1.49$  in $^{197}$Au, indicating a different production mechanism from those in central rapidities.
      
      \par

      \par
\begin{figure}[htb]
      \includegraphics[angle=0,scale=0.42]{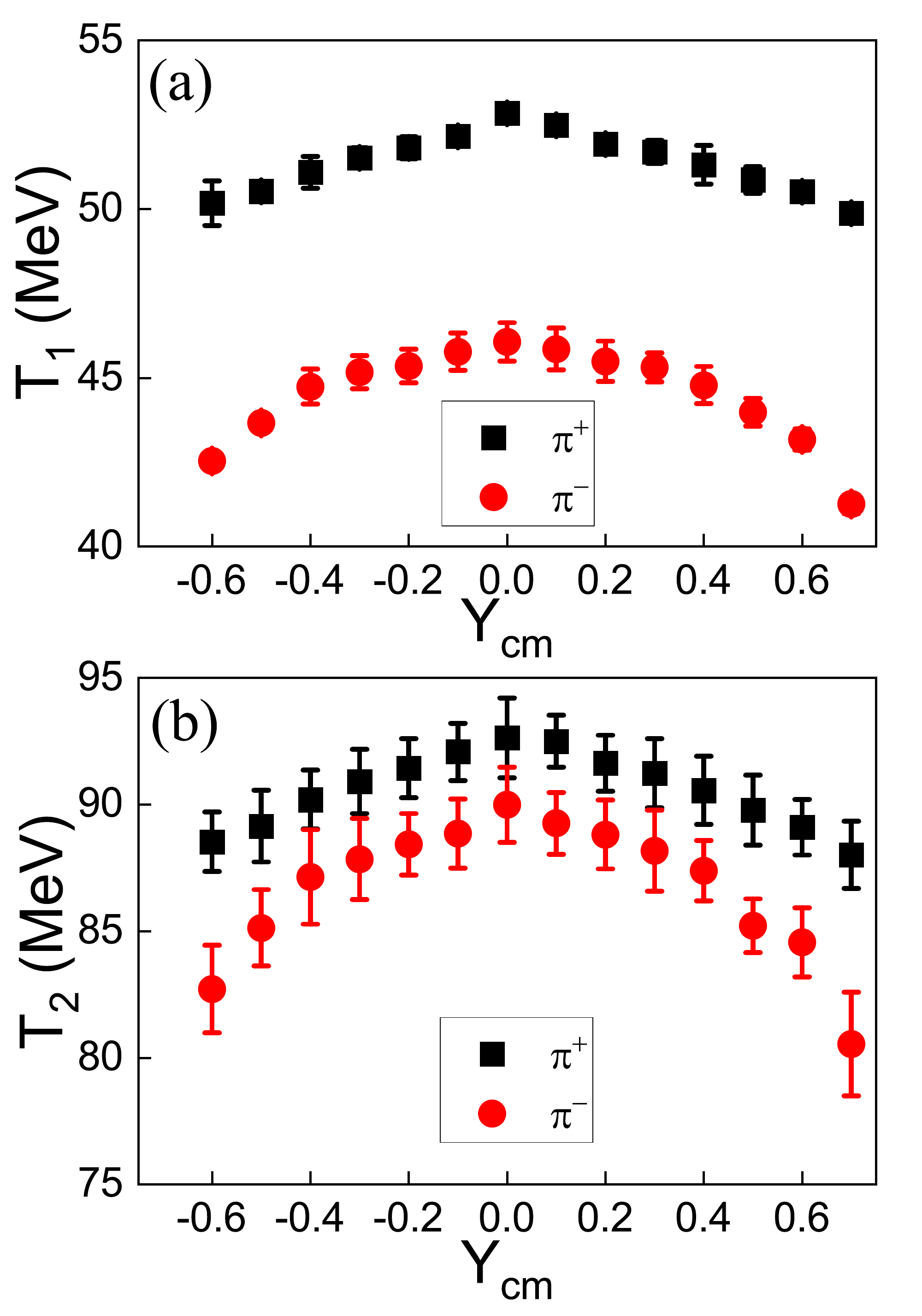}
      \caption{Slope parameters $T_{1}$ (a) and $T_{2}$ (b) of positive (squares) and negative (circles) charged pions for the 0-10\% centrality events  at different rapidity intervals from -0.65 to 0.75.}
      \label{fig3}
      \end{figure}

    Fig.~\ref{fig3} shows the $T_{1}$ and $T_{2}$ from fitting results of charged pions in  0-10\% centrality at different rapidity intervals, while Table~\ref{table1} presents the fit results for mid-rapidity pions at different centrality intervals. $T_{1}$ describes the slope of the low $m_{t}$ part of the spectrum, which contains bulk particles and is usually attributed to pions originating from $\Delta$ decays~\cite{harabasz2020statistical}. $T_{2}$ stands for the slope at higher $m_{t}$, which is often interpreted as a thermal component~\cite{Hong:1997ka}. 
    In an earlier study with the Boltzmann-Uehling-Ulenbeck equation \cite{BALi2}, they proposed that the two-temperature shape of negative pion spectra observed in heavy-ion collisions with beam energies of $\simeq$ 1 GeV/A is due to the different contributions of $\Delta$-resonances produced early and late during the course of the heavy-ion reaction.
     As seen from Fig.~\ref{fig3} and Table ~\ref{table1}, with increasing rapidity and centrality, $T_{1}$ and $T_{2}$ decrease gradually. This can be understood from the much hot source for $\pi$ emission is  created at mid-rapidity as well as in more central collisions, than that for larger centralities and rapidities. Table ~\ref{table1} shows perfect agreements between our IQMD calculations and the HADES data for both $T_1$ and $T_2$. As can be seen from the table~\ref{table1}, there are certain differences between $\pi^{+}$ and $\pi^{-}$. It can be mainly attributed by the difference between positive and negative charged pions to the Coulomb potential. Refs.~\cite{HADES,PhysRevC.57.2536} gave that the Coulomb effect on the transverse momentum of positive and negative charged pions is different. The former are accelerated resulting in  reduced yields at low momenta, while the latter are decelerated resulting in  increased yields at low momenta. Therefore the effect leads to an ``apparent" temperature difference between $\pi^{+}$ and $\pi^{-}$ spectra.  Actually, we checked in our model calculations by switching off the Coulomb potential, the significant difference of the ``apparent" temperature between the positive and negative charged pions disappears, especially for $T_1$.
      
      Here we  would also give some remarks on the pion transverse mass spectra. Even though we took the two-parameter fitting method which is the same as the experimental analysis and the excellent agreements for two ``apparent" temperatures were  reached between the model and the data, the ``apparent" temperatures are not really temperatures, which are in fact the additive results of kinetic freeze-out temperature, Coulomb  interaction as well as the radial flow etc. In principal, the spectra can be fitted by the blast-wave formula which includes the kinetic freeze-out temperature and radial flow parameters. As shown in our previous work \cite{Lv2}, such freeze-out temperatures will be much lower than the ``apparent" temperature extracted from the slope fits. In other words, the radial flow induces the transverse mass spectra harder and then gives larger ``apparent" temperature. But this is beyond our present study since we are trying to compare with the HADES data quantitatively.

\begin{table}[]
\scriptsize
\centering
\caption{The fitting results of two inverse slope parameters $T^{IQMD}_{1}$ and $T^{IQMD}_{2}$ for $\pi ^{+}$ and $\pi ^{-}$ with mid-rapidity cut at four  centrality classes with 10\% bin width and 0-40\% centrality, as well as the comparison with the HADES experimental result from  Ref.~\cite{HADES}.}
\label{table1}
\begin{tabular}{c|cccc}
\hline
$\pi^{+} (\%)$ & $T^{HADES}_{1}$ & $T^{IQMD}_{1}$ & $T^{HADES}_{2}$ & $T^{IQMD}_{2}$ (MeV) \\
\hline
0-10 & 54 & 52.83 & 92 & 92.63\\
10-20 & 51 & 51.50 & 89 & 89.63\\
20-30 & 49 & 49.65 & 86 & 86.90\\
30-40 & 47 & 47.59 & 83 & 83.86\\
0-40 & 52 & 51.06 & 88 & 88.33\\
\hline
$\pi^{-} (\%)$ & $T^{HADES}_{1}$ & $T^{IQMD}_{1}$ & $T^{HADES}_{2}$ & $T^{IQMD}_{2}$ (MeV) \\
\hline
0-10 & 46 & 46.06 & 91 & 89.96\\
10-20 & 43 & 43.64 & 85 & 84.17\\
20-30 & 42 & 40.15 & 82 & 81.71\\
30-40 & 41 & 38.25 & 79 & 78.21\\
0-40 & 44 & 44.28 & 87 & 86.83\\
\hline
\end{tabular}
\end{table}

      \subsection{The three-dimensional HBT radius}
      \par
     For the final-state particle information, which is needed for the HBT study, it is obtained from the IQMD model by  correcting it with the event plane method. The n-$th$ order event plane angle $\Psi _{EP}^{(n)}$ can be defined by the event flow vector $Q_{n,x}$ and  $Q_{n,y}$ as~\cite{PhysRevC.58.1671,PhysRevC.88.014904}
      \begin{eqnarray}
	 \begin{aligned}
      &\Psi _{EP}^{(n)} = \frac{1}{n}\tan^{-1}\left ( \frac{Q_{n,y}}{Q_{n,x}} \right ), \\
      &Q_{n,x}= \sum_{i}^{}\omega_{i}\cos(n\Phi_{i}),\quad Q_{n,y}= \sum_{i}^{}\omega_{i}\sin(n\Phi_{i}),
      \label{eq:ep}
       \end{aligned}
      \end{eqnarray}
where $\Phi_{i}$ and $\omega _{i}$ are the azimuthal angle of the momentum and the weight for the $i$-th particle, respectively. In this work, we chose transverse momentum as weight and construct the first-order ($n$=1) event plane correction.
      \par
\begin{figure}[htb]
      \includegraphics[angle=0,scale=0.35]{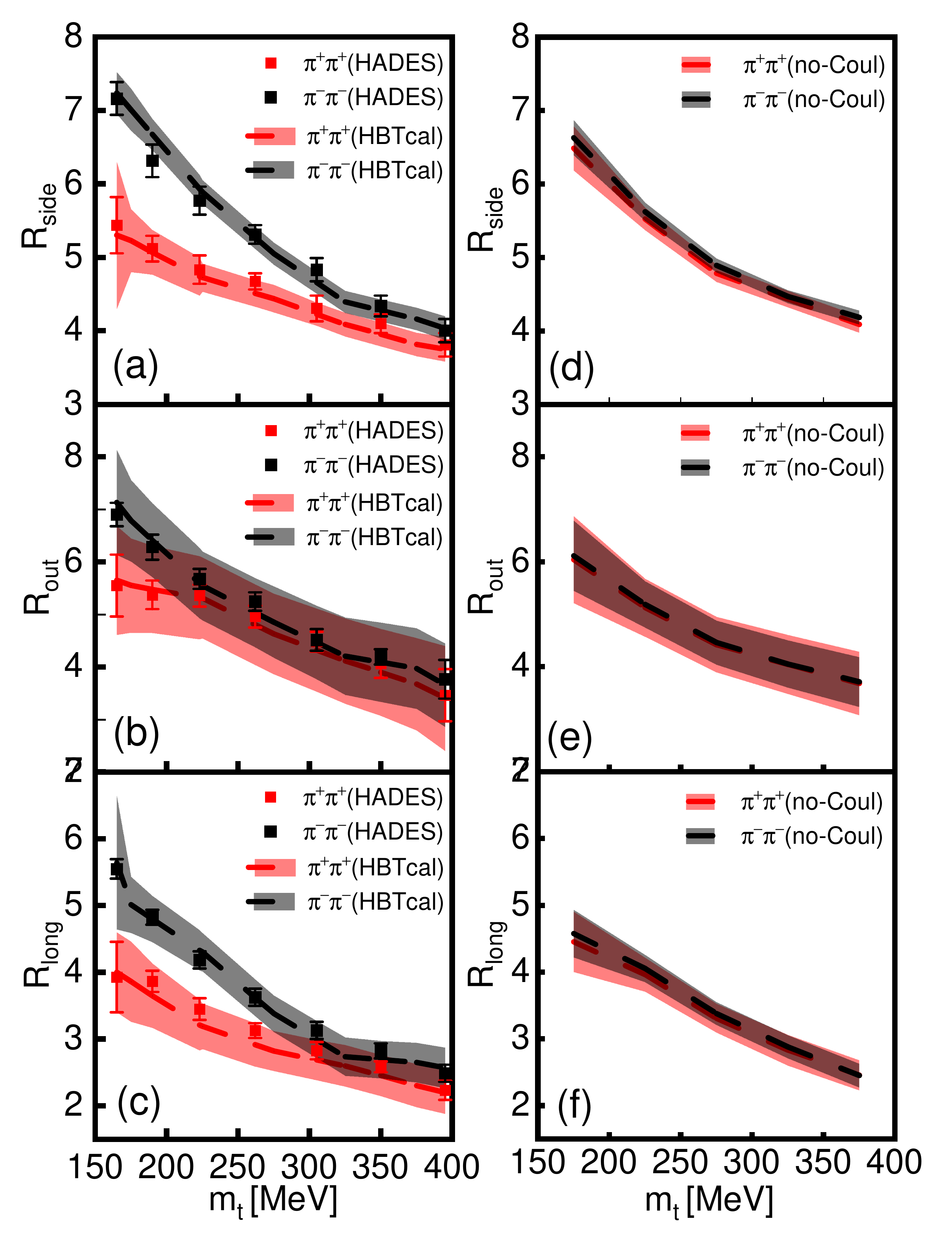}
      \caption{Comparison of three-dimensional HBT radii of charged pion-pairs from the IQMD simulation and the experimental results from the HADES~\cite{ref56}(a,b,c) and the three-dimensional HBT radii of charged pion-pairs from the IQMD model with no-Coulomb interaction(d,e,f). The pion-pairs are selected with mean transverse momentum of 100 - 800 MeV/c and centrality of $0-10\%$. The solid symbols and error bars represent the HADES experimental data, the dotted lines and error bands represent the IQMD simulation results, respectively.}
      \label{fig4}
      \end{figure}

By simplifying Eq.~(\ref{eq:R2_2Phi}), we can obtain expressions for the three-dimensional HBT radii  as~\cite{ref58}:
      \begin{eqnarray}
	 \begin{aligned}
&R_{s}^{2}(K_{\perp },\Phi ,Y) = \left \langle (y \cos\Phi  - x \sin\Phi )^{2} \right \rangle-\left \langle y \cos\Phi  - x \sin\Phi  \right \rangle^{2}\\
&R_{o}^{2}(K_{\perp },\Phi ,Y) = \left \langle (x \cos\Phi  + y \sin\Phi - \beta _{t}t)^{2} \right \rangle \\
&\qquad\qquad\qquad\quad-\left \langle x \cos\Phi  + y \sin\Phi - \beta _{t}t \right \rangle^{2}\\
&R_{l}^{2}(K_{\perp },\Phi ,Y) = \left \langle (z - \beta _{l}t )^{2} \right \rangle-\left \langle z - \beta _{l}t   \right \rangle^{2},
      \label{eq:R_phi}
       \end{aligned}
      \end{eqnarray}
     where the $x$, $y$, $z$, $t$ are the coordinates and emission  times of the %negative 
     charged pions, respectively. $\beta_{\perp}$ and $\beta_{l}$ are the component of the pair velocity in the transverse and longitudinal directions, respectively. 
     To get the $K_{\perp}$($m_{t}$) dependence of HBT radius, the HBT radius in equation~(\ref{eq:R_phi}) is averaged over $\Phi$ and $Y$, which correspond to the zeroth-order HBT radius in Ref.~\cite{retiere2004observable}. The $K_{\perp}$($m_{t}$) dependence of HBT radius is shown in Fig.~\ref{fig4} and is consistent with the data from the HADES Collaboration~\cite{ref56}.

      As we can see from (a,b,c) of Fig.~\ref{fig4}, $R_{side}, R_{long}$, and $R_{out}$ decrease with increasing transverse mass, which are in good agreement with the HADES experimental results. While for low transverse mass, there is an increase/decrease of the HBT radius for
      $\pi^{-}\pi^{-}$/$\pi^{+}\pi^{+}$ pairs, which can be attributed to the Coulomb interactions. And the effect apparently fades away at large transverse momentum~\cite{barz1999combined}. Compared with (d,e,f) of Fig.~\ref{fig4}, after removing the Coulomb reaction, the three-dimensional HBT radii of $\pi^{+}\pi^{+}$ and $\pi^{-}\pi^{-}$ have little difference, which indicates that the increase/decrease excursion of the three-dimensional HBT radius of negative/positive charged pions is indeed due to the Coulomb interaction. And this is consistent with the different ``apparent" temperatures of positive and negative pions as shown in Fig.~\ref{fig3} where in lower momentum region there are more negative pions than positive poins. This can be attributed to the fact that in the reaction system positive pions are always pushed away from the system and the negative pions are trapped in the system via the Coulomb interaction, which results in earlier freeze-out status for positive pions than negative poins.
      The values of $R_{side}$ and $R_{out}$ are essentially the same, but different from $R_{long}$, which is consistent with the definition of the $OSL$ coordinate system and the beam direction. With the increase in $m_{t}$,  $R_{side}$, $R_{out}$ and $R_{long}$ all drop with the transverse mass of pions. This is due to the influence of radial and longitudinal flows, which causes the HBT radius  decreasing gradually with $m_{t}$. The results are consistent with the conclusions in Refs.~\cite{retiere2004observable,barz1999combined}.
      In addition, the values of $R_{side}$ and $R_{out}$ are  bigger than that of $R_{long}$,  indicating  that the expansion effect of the particle source is larger in the sideward and outward directions.

      \begin{figure}[htb]
      \includegraphics[angle=0,scale=0.45]{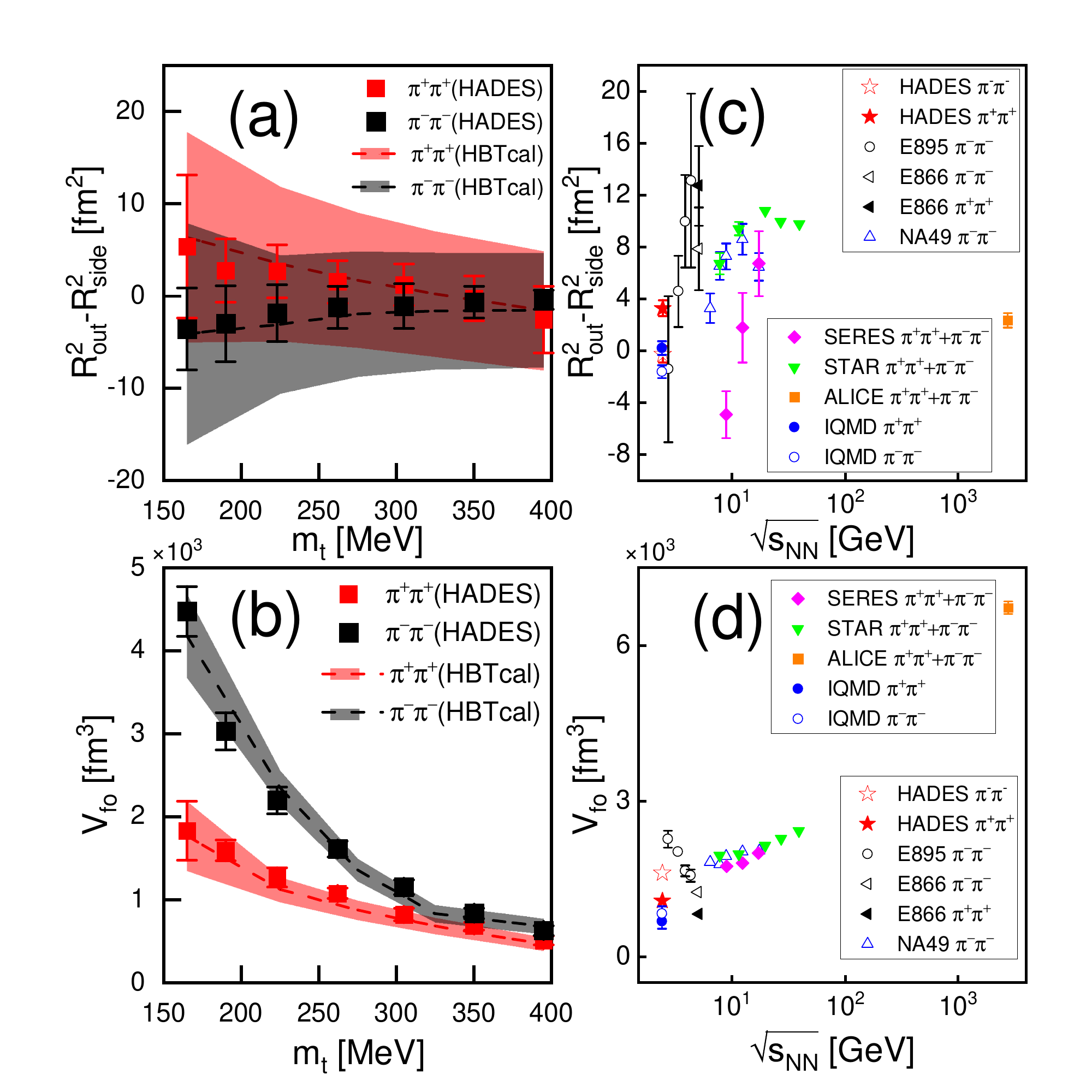}
      \caption{Values of $R_{out}^{2}-R_{side}^{2}$ (top panel) and  $V_{fo}$ (bottom panel) as functions of $m_t$ (left column) and $\sqrt{s_{NN}}$ (right column). The left column are results of charged pion pairs from the IQMD model calculation with mean transverse momentum of 100-800 MeV/c and 0-10\% centrality. The right column shows the comparison of the IQMD simulation results with some experimental data from Ref.~\cite{ref7}.}
      \label{fig5}
      \end{figure}
  
       \begin{figure}[htb]
      \includegraphics[angle=0,scale=0.4]{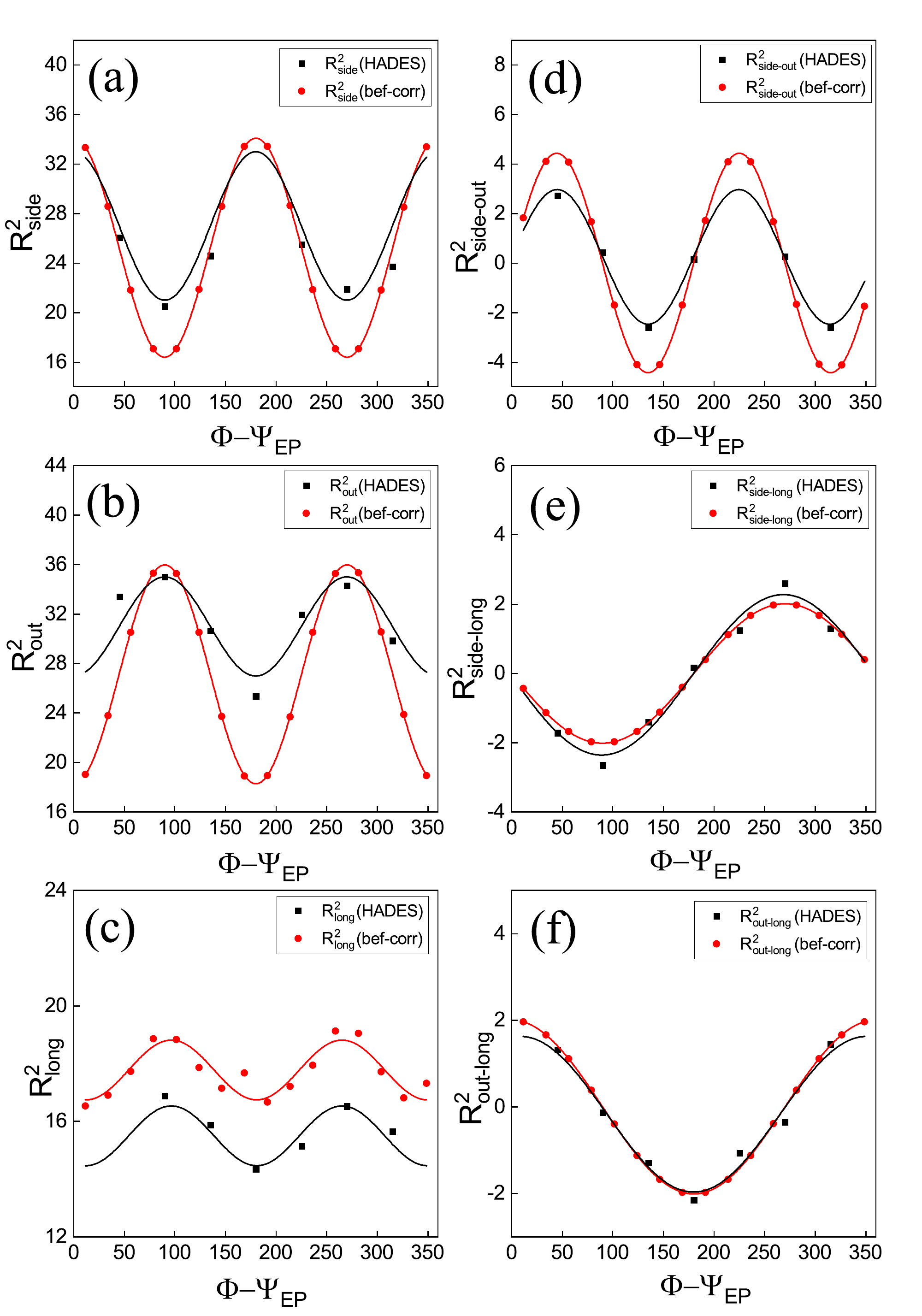}
      \caption{Azimuth dependence of the squared three-dimensional HBT radii for $\pi^{-}$ from the IQMD model with transverse momentum of 150-200 MeV/c in 10-30\% centrality. The left column from top to bottom gives the ``side", ``out" and ``long" radius, respectively,  and the right column from top to bottom shows the ``side-out", ``side-long" and ``out-long" components, respectively. The black square shows the fitting results from the HADES experiment in Ref.~\cite{ref56}, the red circle plots the calculated results from the IQMD model. The data points are fitted to $R(\Phi)$ with Eq.~(\ref{eq:R2_KT_Phi_Y}).}
      \label{fig6}
      \end{figure}
 
       \begin{figure}[htb]
      \includegraphics[angle=0,scale=0.4]{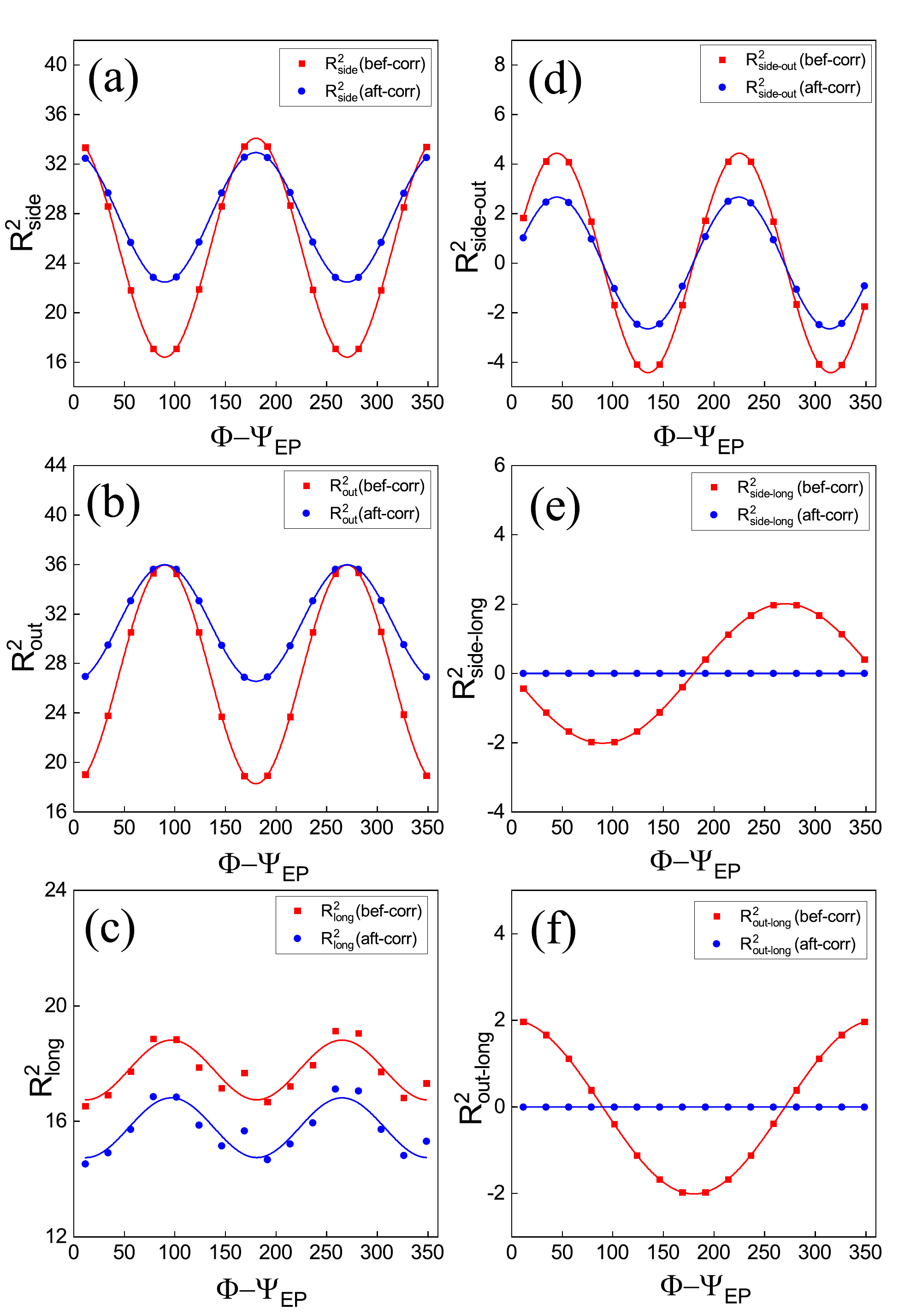}
      \caption{Azimuth dependence of the squared three-dimensional HBT radii, before and after the rotation correction for the reaction plane, for $\pi^{-}$ from the IQMD model with transverse momentum of 150-200 MeV/c and 10-30\% centrality. The left column from top to bottom gives the ``side", ``out" and ``long" radius, respectively,  and the right column from top to bottom shows the ``side-out", ``side-long" and ``out-long" components, respectively. The red circle plots the calculated results before the rotation correction, and the blue triangle displays the calculated results after the rotation correction. The data points are fitted to $R(\Phi)$ with Eq.~(\ref{eq:R2_KT_Phi_Y}).}
      \label{fig7}
      \end{figure}

     $R_{out}^{2}$ and $R_{side}^{2}$ are related to the emission time duration~\cite{ref72,ref73,ref45,ref75,ref76} by the equation $(\Delta\tau)^{2}\approx(R_{out}^{2}-R_{side}^{2})/\langle\beta_{t}^{2}\rangle$~\cite{ref77,ref78}. For the freeze-out volume, it can be expressed as $V_{fo} = (2\pi)^{3/2}R_{side}^{2}R_{long}$. The $m_t$ and $\sqrt{s_{NN}}$ dependences  of $R_{out}^{2}-R_{side}^{2}$ and $V_{fo}$ are shown in Fig.~\ref{fig5}.
      
      As seen from Fig.~\ref{fig5}(a),  the value of $R_{out}^{2}-R_{side}^{2}$ drops as $m_t$ increases and finally turns negative for $m_t>300MeV/c$, and its absolute value increases gradually. This indicates that the emission time duration generally becomes shorter for higher $m_t$ pions.  However, for much higher $m_t$ pions, the value of $R_{out}^{2}-R_{side}^{2}$ becomes negative, which is related to the opacity of the source. The opaqueness of the source affects $R_{out}^{2}-R_{side}^{2}$, and can cause it  negative, thus compensating for the positive contribution of the emission duration. This is consistent with the conclusions in  Ref.\cite{ref56,barz1999combined}. For the value of $V_{fo}$ as shown in Fig.~\ref{fig5}(b), it decreases as $m_t$ increases, indicating the expansion of the emission source since higher $m_t$ pions are emitted earlier.  Moreover, both $R_{out}^{2}-R_{side}^{2}$ and $V_{fo}$ from the IQMD simulation are basically consistent with the trend of measured results from various experiments as a function of $\sqrt{s_{NN}}$ as demonstrated in Fig.~\ref{fig5}(c) and (d). Most measured results in these experiments are in the high energy region. For experiments near the low energy region, the errors are, however, large.

      \subsection{Azimuthal-dependent  HBT radius}
      
      In the current model calculation and experimental measurement, the phase space of the final state particles is asymmetric with respect to the reaction plane in non-central collisions. To correct the particle phase space, we choose the matrix correction method with a so-called S-matrix constructed by $\pi^{-}\pi^{-}$ pairs.
      
      With the $\pi^{-}\pi^{-}$ pairs of mean transverse momentum of 150-200 MeV/c in 10-30\% centrality from the IQMD model, from Eq.~(\ref{eq:smx}), we can get the uncorrected matrix $S$
      \begin{equation}
      S = \begin{bmatrix} 24.7525 & 0.2116 & -0.1877 & -0.6210 \\ 0.2116 & 16.1752 & 0.0817 & 2.0240 \\ -0.1877 & 0.0817 & 31.3588 & -0.0042 \\ -0.6210 & 2.0240 & -0.0042 & 18.4952 \end{bmatrix}
      \end{equation}
      As expected from the consideration of symmetry, only the diagonal elements and $S_{13}$ are significantly different from 0, resulting in six square radii that do not disappear as indicated by Eq.~(\ref{eq:Sjuzhen}). Using the $S$ matrix, we can calculate the HBT radius with Eq.~(\ref{eq:R2_2Phi}), and the results obtained by the IQMD model are compared with experimental data from the HADES Collaboration~\cite{ref56} are shown in Fig.~\ref{fig6}.
    It shows that the HBT radii from the IQMD model and the HADES data have the same phase position and periodicity, and the $R_{sl}$ and $R_{ol}$ obtained by IQMD are in good agreement with those  obtained from HADES. For other HBT radii, the IQMD results showed, however, a larger amplitude than that of the HADES data.
      \par
      The spatial tilt angle of the reaction plane can be calculated from the $S$ matrix,
      \begin{equation}
      \theta_{s} = \frac{1}{2}\tan^{-1}(\frac{2S_{13}}{S_{33}-S_{11}}) = 30.09^{\circ}.
      \end{equation}
     
      Using the corresponding rotation matrix $G_{y}(\theta_{s})$, we rotate the matrix $S$ by $-\theta_s$ around the $y$-axis to obtain the rotation corrected diagonal tensor matrix
%      \begin{scriptsize}
      \begin{eqnarray}
      S = \begin{bmatrix} 24.7525 & 0.2116 & -0.1877 & -0.6210 \\ 0.2116 & 18.0016 & 0.0817 & 0.0017 \\ -0.1877 & 0.0817 & 31.3588 & -0.0042 \\ -0.6210 & 0.0017 & -0.0042 & 16.6682 \end{bmatrix}.
      \end{eqnarray}
%      \end{scriptsize}
  
      The results of azimuth dependent HBT radius calculated before and after the rotation correction are shown in Fig.~\ref{fig7}, respectively.
      As seen from Fig.~\ref{fig7}, for the first-order HBT radius, the oscillation structure of the corrected HBT radius remains unchanged and the amplitude decreases. For the second-order HBT radius, $R_{sl}$ and $R_{ol}$ are approximately zero after correction. The most immediate reason is that the value of $S_{13}$ becomes 0 after the correction, so that the value of $R^{2}_{ol}$ and $R^{2}_{sl}$ in Eq.~(\ref{eq:R2_2Phi}) is approximately equal to 0. In addition, it is indicated in Refs.~\cite{ref8,LISA20001} that the cross terms $R^{2}_{ol}$ and $R^{2}_{sl}$ cancel to 0 due to the high symmetry of the source after correction. This result is also coincident with that from the RQMD model~\cite{ref58}.

      \par
      Finally, we discuss the eccentricities of $xy$- and $zy$-plane for  emissions of pion-pairs in the viewpoint of the eigenvalues of matrix $S$, namely the temporal and geometrical variances $\sigma _{t}^{2}$, $\sigma _{x}^{2}$, $\sigma _{y}^{2}$ and $\sigma _{z}^{2}$. Using these non-zero eigenvalues, the eccentricities of $xy$- and $zy$-plane of pion-pairs can be calculated  as follows~\cite{ref56}:
      \begin{equation}
      \varepsilon _{xy} = \frac{\sigma _{y}^{2}-\sigma _{x}^{2}}{\sigma _{y}^{2}+\sigma _{x}^{2}},\qquad \varepsilon _{zy} = \frac{\sigma _{y}^{2}-\sigma _{z}^{2}}{\sigma _{y}^{2}+\sigma _{z}^{2}}.
      \end{equation}

      \begin{figure}[htb]
      \includegraphics[angle=0,scale=0.6]{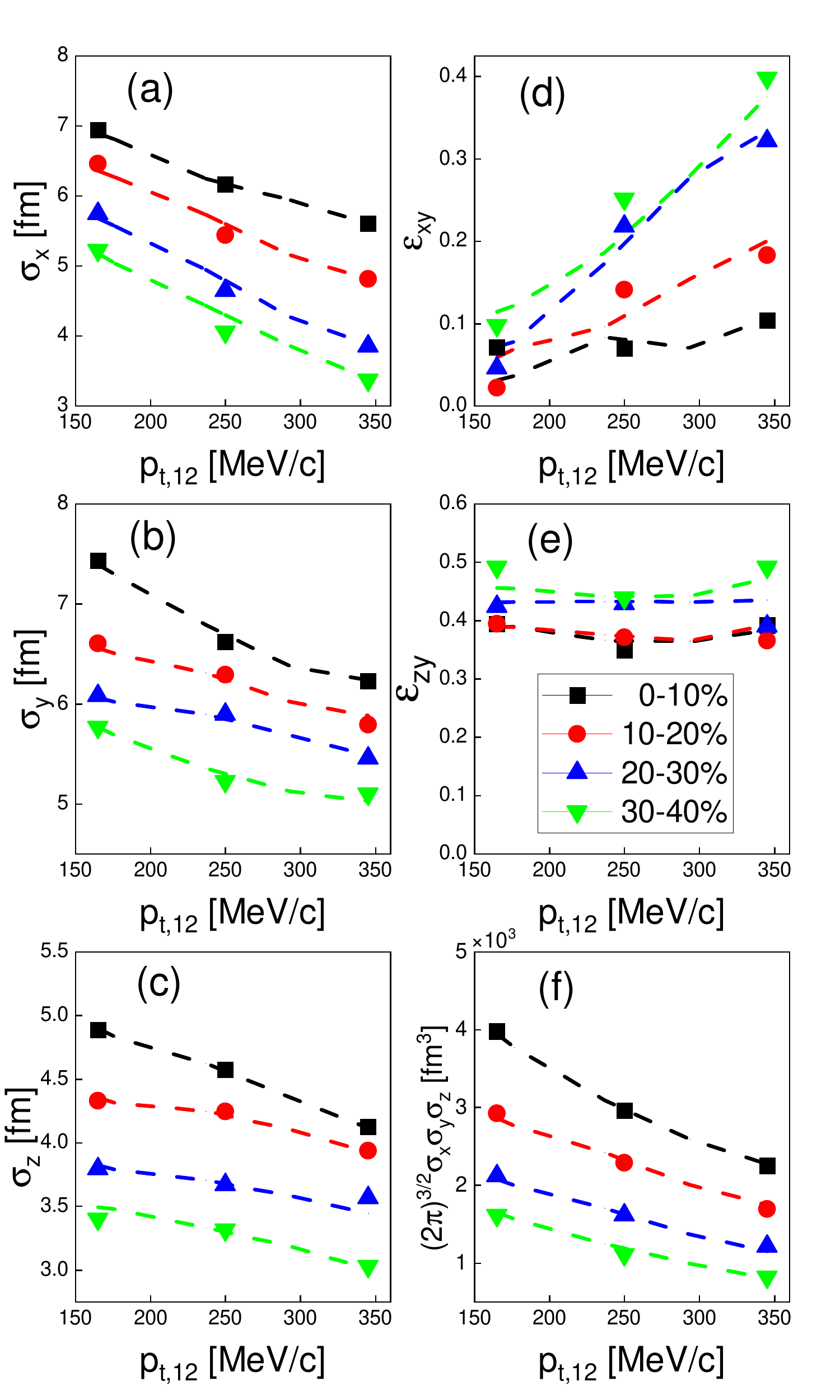}
      \caption{The eccentric distances  $\sigma _{x}$ (a), $\sigma _{y}$ (b) and $\sigma _{z}$ (c), the $xy$-eccentricity (d), the $zy$-eccentricity (e) and the freeze-out volume (f) of $\pi^{-}\pi^{-}$ pairs as a function of pair transverse momentum for different centrality classes. The solid symbols and lines represent the fitting result from the HADES experiment in Ref.~\cite{ref56} and our IQMD simulation results, respectively.}
      \label{fig8}
      \end{figure}

      \par
       Fig.~\ref{fig8} shows the $\pi^{-}\pi^{-}$ pair transverse-momentum ($p_{t,12}$) dependence of three dimensional eccentric distances ($\sigma_x$, $\sigma_y$ and $\sigma_z$) as well as the $xy$- and $zy$- eccentricities ($\epsilon_{xy}$ and $\epsilon_{zy}$) and the freeze-out volume by an approximate formula of $V_{f} = (2\pi)^{3/2}\sigma_{x}\sigma_{y}\sigma_{z}$ for different collision centralities. We can see that all $\sigma_{x}$, $\sigma_{y}$ and $\sigma_{z}$ decrease with increasing pair transverse momentum as well as centrality, which indicates that $\pi^{-}\pi^{-}$ pairs with higher pair transverse momentum and/or in less central collisions have less dispersion  in the spatial distribution, which is also consistent with the trend of HBT radii with $m_t$. We can also see that the $xy$-eccentricity increases with the pair transverse momentum, indicating a more asymmetric transverse emission for pions with higher transverse momentum. There is, however, no obvious  $p_{t,12}$ dependence for the $zy$-eccentricity. For more central collisions, the eccentricity of $xy$-plane becomes very small, approaching essentially zero. This means that the distribution of the final-state particles in the $xy$- plane is close to a circular shape. With the increase in centrality, the degree of deviation from the circular-like shape in the $xy$- and $zy$-plane increases gradually. The deduced freeze-out $V_{f}$ decreases with the $p_{t,12}$ and centrality, which means that $\pi^{-}\pi^{-}$ pairs with higher pair transverse momentum and/or in off-central  collisions have a smaller volume in their homogeneity region. Our  calculated results are in good agreement with the experimental results from the  HADES data~\cite{ref56}.

      \section{Summary}
      \label{sec:summary}
      \par
      In summary,  the yield of charged pions and their three-dimensional HBT radii were calculated by a simulation of Au + Au collision at 1.23$A$ GeV with the IQMD model. Even though there is an overpredication of single $\pi$ spectra in comparison with the data, the $\pi^-/\pi^+$ and ``apparent" temperature parameters match up with the data. The $R_{out}^{2}-R_{side}^{2}$ and $V_{fo}$ obtained by our calculations are basically consistent with the observed results given in various experiments. We also calculated the azimuthal dependence of HBT radii by the matrix-element method and corrected it by the rotation matrix. In this method, the $\Phi$-dependent HBT radii can be calculated from five components of the spatial correlation tensor $S_{\mu\nu}$, and corrected by the rotation matrix. In addition, we also calculated the $xy$- and $zy$-plane eccentricities as a function of pair transverse momentum and centrality. The results show that the $xy$-eccentricity of the pion emission source is approximately zero in the most central collision, which means an approximately circular shape in the $xy$-plane. With the increase in pair transverse momentum, the $xy$-eccentricity increases gradually, indicating a more asymmetric transverse emission source for higher pair transverse momentum. However, there is no obvious $p_{t,12}$ dependence in the $zy$-eccentricity. The deduced freeze-out volume $V_{fo}$ decreases with the increasing  $p_{t,12}$ and centrality, which means that $\pi^{-}\pi^{-}$ pairs with higher pair transverse momentum and in off-central collision have a smaller homogeneity region volume.

      \begin{acknowledgements}
      This work was supported in part by the National Natural Science Foundation of China under contract Nos. 11890714, 11875066, 11421505, 11775288, and and 12147101, the National Key R\&D Program of China under Grant Nos. 2016YFE0100900 and 2018YFE0104600, and by Guangdong Major Project of Basic and Applied Basic Research No. 2020B0301030008.
      \end{acknowledgements}

      \end{CJK*}

      \bibliography{myref}

\begin{thebibliography}{83}%
\makeatletter
\providecommand \@ifxundefined [1]{%
 \@ifx{#1\undefined}
}%
\providecommand \@ifnum [1]{%
 \ifnum #1\expandafter \@firstoftwo
 \else \expandafter \@secondoftwo
 \fi
}%
\providecommand \@ifx [1]{%
 \ifx #1\expandafter \@firstoftwo
 \else \expandafter \@secondoftwo
 \fi
}%
\providecommand \natexlab [1]{#1}%
\providecommand \enquote  [1]{``#1''}%
\providecommand \bibnamefont  [1]{#1}%
\providecommand \bibfnamefont [1]{#1}%
\providecommand \citenamefont [1]{#1}%
\providecommand \href@noop [0]{\@secondoftwo}%
\providecommand \href [0]{\begingroup \@sanitize@url \@href}%
\providecommand \@href[1]{\@@startlink{#1}\@@href}%
\providecommand \@@href[1]{\endgroup#1\@@endlink}%
\providecommand \@sanitize@url [0]{\catcode `\\12\catcode `\$12\catcode
  `\&12\catcode `\#12\catcode `\^12\catcode `\_12\catcode `\%12\relax}%
\providecommand \@@startlink[1]{}%
\providecommand \@@endlink[0]{}%
\providecommand \url  [0]{\begingroup\@sanitize@url \@url }%
\providecommand \@url [1]{\endgroup\@href {#1}{\urlprefix }}%
\providecommand \urlprefix  [0]{URL }%
\providecommand \Eprint [0]{\href }%
\providecommand \doibase [0]{http://dx.doi.org/}%
\providecommand \selectlanguage [0]{\@gobble}%
\providecommand \bibinfo  [0]{\@secondoftwo}%
\providecommand \bibfield  [0]{\@secondoftwo}%
\providecommand \translation [1]{[#1]}%
\providecommand \BibitemOpen [0]{}%
\providecommand \bibitemStop [0]{}%
\providecommand \bibitemNoStop [0]{.\EOS\space}%
\providecommand \EOS [0]{\spacefactor3000\relax}%
\providecommand \BibitemShut  [1]{\csname bibitem#1\endcsname}%
\let\auto@bib@innerbib\@empty
%</preamble>
\bibitem [{\citenamefont {Hanbury~Brown}\ and\ \citenamefont
  {Twiss}(1956)}]{ref1}%
  \BibitemOpen
  \bibfield  {author} {\bibinfo {author} {\bibfnamefont {R.}~\bibnamefont
  {Hanbury~Brown}}\ and\ \bibinfo {author} {\bibfnamefont {R.~Q.}\ \bibnamefont
  {Twiss}},\ }\href {\doibase 10.1038/1781046a0} {\bibfield  {journal}
  {\bibinfo  {journal} {Nature}\ }\textbf {\bibinfo {volume} {178}},\ \bibinfo
  {pages} {1046} (\bibinfo {year} {1956})}\BibitemShut {NoStop}%
\bibitem [{\citenamefont {Alver}\ and\ \citenamefont {Roland}(2010)}]{ref2}%
  \BibitemOpen
  \bibfield  {author} {\bibinfo {author} {\bibfnamefont {B.}~\bibnamefont
  {Alver}}\ and\ \bibinfo {author} {\bibfnamefont {G.}~\bibnamefont {Roland}},\
  }\href {\doibase 10.1103/PhysRevC.82.039903} {\bibfield  {journal} {\bibinfo
  {journal} {Phys. Rev. C}\ }\textbf {\bibinfo {volume} {81}},\ \bibinfo
  {pages} {054905} (\bibinfo {year} {2010})},\ \bibinfo {note} {[Erratum: Phys.
  Rev. C 82, 039903 (2010)]},\ \Eprint {http://arxiv.org/abs/1003.0194}
  {arXiv:1003.0194 [nucl-th]} \BibitemShut {NoStop}%
\bibitem [{\citenamefont {Lisa}\ \emph {et~al.}(2005)\citenamefont {Lisa},
  \citenamefont {Pratt}, \citenamefont {Soltz},\ and\ \citenamefont
  {Wiedemann}}]{ref3}%
  \BibitemOpen
  \bibfield  {author} {\bibinfo {author} {\bibfnamefont {M.~A.}\ \bibnamefont
  {Lisa}}, \bibinfo {author} {\bibfnamefont {S.}~\bibnamefont {Pratt}},
  \bibinfo {author} {\bibfnamefont {R.}~\bibnamefont {Soltz}}, \ and\ \bibinfo
  {author} {\bibfnamefont {U.}~\bibnamefont {Wiedemann}},\ }\href {\doibase
  10.1146/annurev.nucl.55.090704.151533} {\bibfield  {journal} {\bibinfo
  {journal} {Ann. Rev. Nucl. Part. Sci.}\ }\textbf {\bibinfo {volume} {55}},\
  \bibinfo {pages} {357} (\bibinfo {year} {2005})},\ \Eprint
  {http://arxiv.org/abs/nucl-ex/0505014} {arXiv:nucl-ex/0505014} \BibitemShut
  {NoStop}%
\bibitem [{\citenamefont {Lin}\ \emph {et~al.}(2005)\citenamefont {Lin},
  \citenamefont {Ko}, \citenamefont {Li}, \citenamefont {Zhang},\ and\
  \citenamefont {Pal}}]{ref4}%
  \BibitemOpen
  \bibfield  {author} {\bibinfo {author} {\bibfnamefont {Z.-W.}\ \bibnamefont
  {Lin}}, \bibinfo {author} {\bibfnamefont {C.~M.}\ \bibnamefont {Ko}},
  \bibinfo {author} {\bibfnamefont {B.-A.}\ \bibnamefont {Li}}, \bibinfo
  {author} {\bibfnamefont {B.}~\bibnamefont {Zhang}}, \ and\ \bibinfo {author}
  {\bibfnamefont {S.}~\bibnamefont {Pal}},\ }\href {\doibase
  10.1103/PhysRevC.72.064901} {\bibfield  {journal} {\bibinfo  {journal} {Phys.
  Rev. C}\ }\textbf {\bibinfo {volume} {72}},\ \bibinfo {pages} {064901}
  (\bibinfo {year} {2005})},\ \Eprint {http://arxiv.org/abs/nucl-th/0411110}
  {arXiv:nucl-th/0411110} \BibitemShut {NoStop}%
\bibitem [{\citenamefont {Akkelin}\ and\ \citenamefont
  {Sinyukov}(1995)}]{ref5}%
  \BibitemOpen
  \bibfield  {author} {\bibinfo {author} {\bibfnamefont {S.~V.}\ \bibnamefont
  {Akkelin}}\ and\ \bibinfo {author} {\bibfnamefont {Y.~M.}\ \bibnamefont
  {Sinyukov}},\ }\href {\doibase 10.1016/0370-2693(95)00765-D} {\bibfield
  {journal} {\bibinfo  {journal} {Phys. Lett. B}\ }\textbf {\bibinfo {volume}
  {356}},\ \bibinfo {pages} {525} (\bibinfo {year} {1995})}\BibitemShut
  {NoStop}%
\bibitem [{\citenamefont {Goldhaber}\ \emph {et~al.}(1960)\citenamefont
  {Goldhaber}, \citenamefont {Goldhaber}, \citenamefont {Lee},\ and\
  \citenamefont {Pais}}]{ref6}%
  \BibitemOpen
  \bibfield  {author} {\bibinfo {author} {\bibfnamefont {G.}~\bibnamefont
  {Goldhaber}}, \bibinfo {author} {\bibfnamefont {S.}~\bibnamefont
  {Goldhaber}}, \bibinfo {author} {\bibfnamefont {W.-Y.}\ \bibnamefont {Lee}},
  \ and\ \bibinfo {author} {\bibfnamefont {A.}~\bibnamefont {Pais}},\ }\href
  {\doibase 10.1103/PhysRev.120.300} {\bibfield  {journal} {\bibinfo  {journal}
  {Phys. Rev.}\ }\textbf {\bibinfo {volume} {120}},\ \bibinfo {pages} {300}
  (\bibinfo {year} {1960})}\BibitemShut {NoStop}%
\bibitem [{\citenamefont {Adamczewski-Musch}\ \emph {et~al.}(2019)\citenamefont
  {Adamczewski-Musch} \emph {et~al.}}]{ref7}%
  \BibitemOpen
  \bibfield  {author} {\bibinfo {author} {\bibfnamefont {J.}~\bibnamefont
  {Adamczewski-Musch}} \emph {et~al.} (\bibinfo {collaboration} {HADES
  Collaboration}),\ }\href {\doibase 10.1016/j.physletb.2019.06.047} {\bibfield
   {journal} {\bibinfo  {journal} {Phys. Lett. B}\ }\textbf {\bibinfo {volume}
  {795}},\ \bibinfo {pages} {446} (\bibinfo {year} {2019})},\ \Eprint
  {http://arxiv.org/abs/1811.06213} {arXiv:1811.06213 [nucl-ex]} \BibitemShut
  {NoStop}%
\bibitem [{\citenamefont {Wiedemann}(1998)}]{ref8}%
  \BibitemOpen
  \bibfield  {author} {\bibinfo {author} {\bibfnamefont {U.~A.}\ \bibnamefont
  {Wiedemann}},\ }\href {\doibase 10.1103/PhysRevC.57.266} {\bibfield
  {journal} {\bibinfo  {journal} {Phys. Rev. C}\ }\textbf {\bibinfo {volume}
  {57}},\ \bibinfo {pages} {266} (\bibinfo {year} {1998})},\ \Eprint
  {http://arxiv.org/abs/nucl-th/9707046} {arXiv:nucl-th/9707046} \BibitemShut
  {NoStop}%
\bibitem [{\citenamefont {Bowler}(1988)}]{ref9}%
  \BibitemOpen
  \bibfield  {author} {\bibinfo {author} {\bibfnamefont {M.~G.}\ \bibnamefont
  {Bowler}},\ }\href {\doibase 10.1007/BF01560395} {\bibfield  {journal}
  {\bibinfo  {journal} {Z. Phys. C}\ }\textbf {\bibinfo {volume} {39}},\
  \bibinfo {pages} {81} (\bibinfo {year} {1988})}\BibitemShut {NoStop}%
\bibitem [{\citenamefont {Akkelin}\ and\ \citenamefont
  {Sinyukov}(1996)}]{ref10}%
  \BibitemOpen
  \bibfield  {author} {\bibinfo {author} {\bibfnamefont {S.~V.}\ \bibnamefont
  {Akkelin}}\ and\ \bibinfo {author} {\bibfnamefont {Y.~M.}\ \bibnamefont
  {Sinyukov}},\ }\href {\doibase 10.1007/s002880050271} {\bibfield  {journal}
  {\bibinfo  {journal} {Z. Phys. C}\ }\textbf {\bibinfo {volume} {72}},\
  \bibinfo {pages} {501} (\bibinfo {year} {1996})}\BibitemShut {NoStop}%
\bibitem [{\citenamefont {Fang}\ \emph {et~al.}(2016)\citenamefont {Fang},
  \citenamefont {Ma}, \citenamefont {Sun} \emph {et~al.}}]{ref11}%
  \BibitemOpen
  \bibfield  {author} {\bibinfo {author} {\bibfnamefont {D.~Q.}\ \bibnamefont
  {Fang}}, \bibinfo {author} {\bibfnamefont {Y.~G.}\ \bibnamefont {Ma}},
  \bibinfo {author} {\bibfnamefont {X.~Y.}\ \bibnamefont {Sun}},  \emph
  {et~al.},\ }\href {\doibase 10.1103/PhysRevC.94.044621} {\bibfield  {journal}
  {\bibinfo  {journal} {Phys. Rev. C}\ }\textbf {\bibinfo {volume} {94}},\
  \bibinfo {pages} {044621} (\bibinfo {year} {2016})},\ \Eprint
  {http://arxiv.org/abs/1610.04731} {arXiv:1610.04731 [nucl-ex]} \BibitemShut
  {NoStop}%
\bibitem [{\citenamefont {Lin}\ and\ \citenamefont {Ko}(2002)}]{ref12}%
  \BibitemOpen
  \bibfield  {author} {\bibinfo {author} {\bibfnamefont {Z.-w.}\ \bibnamefont
  {Lin}}\ and\ \bibinfo {author} {\bibfnamefont {C.~M.}\ \bibnamefont {Ko}},\
  }\href {\doibase 10.1103/PhysRevC.65.034904} {\bibfield  {journal} {\bibinfo
  {journal} {Phys. Rev. C}\ }\textbf {\bibinfo {volume} {65}},\ \bibinfo
  {pages} {034904} (\bibinfo {year} {2002})},\ \Eprint
  {http://arxiv.org/abs/nucl-th/0108039} {arXiv:nucl-th/0108039} \BibitemShut
  {NoStop}%
\bibitem [{\citenamefont {Yang}\ and\ \citenamefont {Zhang}(2016)}]{ref13}%
  \BibitemOpen
  \bibfield  {author} {\bibinfo {author} {\bibfnamefont {J.}~\bibnamefont
  {Yang}}\ and\ \bibinfo {author} {\bibfnamefont {W.-N.}\ \bibnamefont
  {Zhang}},\ }\href {\doibase 10.1007/s41365-016-0145-z} {\bibfield  {journal}
  {\bibinfo  {journal} {Nucl. Sci. Tech.}\ }\textbf {\bibinfo {volume} {27}},\
  \bibinfo {pages} {147} (\bibinfo {year} {2016})}\BibitemShut {NoStop}%
\bibitem [{\citenamefont {Li}\ \emph {et~al.}(2021)\citenamefont {Li},
  \citenamefont {Ru},\ and\ \citenamefont {Hu}}]{LiLY}%
  \BibitemOpen
  \bibfield  {author} {\bibinfo {author} {\bibfnamefont {L.-Y.}\ \bibnamefont
  {Li}}, \bibinfo {author} {\bibfnamefont {P.}~\bibnamefont {Ru}}, \ and\
  \bibinfo {author} {\bibfnamefont {Y.}~\bibnamefont {Hu}},\ }\href {\doibase
  https://doi.org/10.1007/s41365-021-00853-7} {\bibfield  {journal} {\bibinfo
  {journal} {Nucl. Sci. Tech.}\ }\textbf {\bibinfo {volume} {32}},\ \bibinfo
  {pages} {19} (\bibinfo {year} {2021})}\BibitemShut {NoStop}%
\bibitem [{\citenamefont {Christie}\ \emph {et~al.}(1993)\citenamefont
  {Christie} \emph {et~al.}}]{ref14}%
  \BibitemOpen
  \bibfield  {author} {\bibinfo {author} {\bibfnamefont {W.~B.}\ \bibnamefont
  {Christie}} \emph {et~al.},\ }\href {\doibase 10.1103/PhysRevC.47.779}
  {\bibfield  {journal} {\bibinfo  {journal} {Phys. Rev. C}\ }\textbf {\bibinfo
  {volume} {47}},\ \bibinfo {pages} {779} (\bibinfo {year} {1993})}\BibitemShut
  {NoStop}%
\bibitem [{\citenamefont {Christie}\ \emph {et~al.}(1992)\citenamefont
  {Christie} \emph {et~al.}}]{ref15}%
  \BibitemOpen
  \bibfield  {author} {\bibinfo {author} {\bibfnamefont {W.~B.}\ \bibnamefont
  {Christie}} \emph {et~al.},\ }\href {\doibase 10.1103/PhysRevC.45.2836}
  {\bibfield  {journal} {\bibinfo  {journal} {Phys. Rev. C}\ }\textbf {\bibinfo
  {volume} {45}},\ \bibinfo {pages} {2836} (\bibinfo {year}
  {1992})}\BibitemShut {NoStop}%
\bibitem [{\citenamefont {Jeltes}\ \emph {et~al.}(2007)\citenamefont {Jeltes},
  \citenamefont {Hogervorst} \emph {et~al.}}]{ref_nature_2007}%
  \BibitemOpen
  \bibfield  {author} {\bibinfo {author} {\bibfnamefont {J.~M.}\ \bibnamefont
  {Jeltes}, \bibfnamefont {T.~McNamara}}, \bibinfo {author} {\bibfnamefont
  {W.}~\bibnamefont {Hogervorst}},  \emph {et~al.},\ }\href {\doibase
  10.1038/nature05513} {\bibfield  {journal} {\bibinfo  {journal} {Nature}\
  }\textbf {\bibinfo {volume} {445}},\ \bibinfo {pages} {402} (\bibinfo {year}
  {2007})}\BibitemShut {NoStop}%
\bibitem [{\citenamefont {Adamczyk}\ \emph
  {et~al.}(2015{\natexlab{a}})\citenamefont {Adamczyk}, \citenamefont {Adkins},
  \citenamefont {Agakishievet} \emph {et~al.}}]{ref_nature_star}%
  \BibitemOpen
  \bibfield  {author} {\bibinfo {author} {\bibfnamefont {L.}~\bibnamefont
  {Adamczyk}}, \bibinfo {author} {\bibfnamefont {J.~K.}\ \bibnamefont
  {Adkins}}, \bibinfo {author} {\bibfnamefont {G.}~\bibnamefont
  {Agakishievet}},  \emph {et~al.} (\bibinfo {collaboration} {STAR
  Collaboration}),\ }\href {\doibase 10.1038/nature15724} {\bibfield  {journal}
  {\bibinfo  {journal} {Nature}\ }\textbf {\bibinfo {volume} {527}},\ \bibinfo
  {pages} {345} (\bibinfo {year} {2015}{\natexlab{a}})}\BibitemShut {NoStop}%
\bibitem [{\citenamefont {Adam}\ \emph {et~al.}(2019)\citenamefont {Adam} \emph
  {et~al.}}]{Neha}%
  \BibitemOpen
  \bibfield  {author} {\bibinfo {author} {\bibfnamefont {J.}~\bibnamefont
  {Adam}} \emph {et~al.} (\bibinfo {collaboration} {STAR Collaboration}),\
  }\href {\doibase https://doi.org/10.1016/j.physletb.2019.01.055} {\bibfield
  {journal} {\bibinfo  {journal} {Phys. Lett. B}\ }\textbf {\bibinfo {volume}
  {790}},\ \bibinfo {pages} {490} (\bibinfo {year} {2019})}\BibitemShut
  {NoStop}%
\bibitem [{\citenamefont {Acharya}\ \emph {et~al.}(2020)\citenamefont
  {Acharya}, \citenamefont {Acosta}, \citenamefont {Adam} \emph
  {et~al.}}]{ref_nature_alice}%
  \BibitemOpen
  \bibfield  {author} {\bibinfo {author} {\bibfnamefont {S.}~\bibnamefont
  {Acharya}}, \bibinfo {author} {\bibfnamefont {F.~T.}\ \bibnamefont {Acosta}},
  \bibinfo {author} {\bibfnamefont {J.}~\bibnamefont {Adam}},  \emph {et~al.}
  (\bibinfo {collaboration} {ALICE Collaboration}),\ }\href {\doibase
  10.1038/s41586-020-3001-6} {\bibfield  {journal} {\bibinfo  {journal}
  {Nature}\ }\textbf {\bibinfo {volume} {588}},\ \bibinfo {pages} {232}
  (\bibinfo {year} {2020})}\BibitemShut {NoStop}%
\bibitem [{\citenamefont {Bertsch}\ \emph {et~al.}(1988)\citenamefont
  {Bertsch}, \citenamefont {Gong},\ and\ \citenamefont {Tohyama}}]{ref16}%
  \BibitemOpen
  \bibfield  {author} {\bibinfo {author} {\bibfnamefont {G.}~\bibnamefont
  {Bertsch}}, \bibinfo {author} {\bibfnamefont {M.}~\bibnamefont {Gong}}, \
  and\ \bibinfo {author} {\bibfnamefont {M.}~\bibnamefont {Tohyama}},\ }\href
  {\doibase 10.1103/PhysRevC.37.1896} {\bibfield  {journal} {\bibinfo
  {journal} {Phys. Rev. C}\ }\textbf {\bibinfo {volume} {37}},\ \bibinfo
  {pages} {1896} (\bibinfo {year} {1988})}\BibitemShut {NoStop}%
\bibitem [{\citenamefont {Kapusta}\ and\ \citenamefont {Li}(2004)}]{ref17}%
  \BibitemOpen
  \bibfield  {author} {\bibinfo {author} {\bibfnamefont {J.~I.}\ \bibnamefont
  {Kapusta}}\ and\ \bibinfo {author} {\bibfnamefont {Y.}~\bibnamefont {Li}},\
  }\href {\doibase 10.1088/0954-3899/30/8/060} {\bibfield  {journal} {\bibinfo
  {journal} {J. Phys. G}\ }\textbf {\bibinfo {volume} {30}},\ \bibinfo {pages}
  {S1069} (\bibinfo {year} {2004})}\BibitemShut {NoStop}%
\bibitem [{\citenamefont {Christopher}\ and\ \citenamefont
  {Kapusta}(2017)}]{ref_plum}%
  \BibitemOpen
  \bibfield  {author} {\bibinfo {author} {\bibfnamefont {P.}~\bibnamefont
  {Christopher}}\ and\ \bibinfo {author} {\bibfnamefont {J.~I.}\ \bibnamefont
  {Kapusta}},\ }\href {\doibase https://doi.org/10.1103/PhysRevC.95.044910}
  {\bibfield  {journal} {\bibinfo  {journal} {Phys. Rev. C}\ }\textbf {\bibinfo
  {volume} {95}},\ \bibinfo {pages} {044910} (\bibinfo {year}
  {2017})}\BibitemShut {NoStop}%
\bibitem [{\citenamefont {He}\ \emph {et~al.}(2020)\citenamefont {He},
  \citenamefont {Zhang}, \citenamefont {Ma}, \citenamefont {Chen},\ and\
  \citenamefont {Zhong}}]{ref18}%
  \BibitemOpen
  \bibfield  {author} {\bibinfo {author} {\bibfnamefont {J.}~\bibnamefont
  {He}}, \bibinfo {author} {\bibfnamefont {S.}~\bibnamefont {Zhang}}, \bibinfo
  {author} {\bibfnamefont {Y.-G.}\ \bibnamefont {Ma}}, \bibinfo {author}
  {\bibfnamefont {J.}~\bibnamefont {Chen}}, \ and\ \bibinfo {author}
  {\bibfnamefont {C.}~\bibnamefont {Zhong}},\ }\href {\doibase
  https://doi.org/10.1140/epja/s10050-019-00002-0} {\bibfield  {journal}
  {\bibinfo  {journal} {Eur. Phys. J. A}\ }\textbf {\bibinfo {volume} {56}},\
  \bibinfo {pages} {52} (\bibinfo {year} {2020})}\BibitemShut {NoStop}%
\bibitem [{\citenamefont {Wang}\ \emph {et~al.}(2021)\citenamefont {Wang},
  \citenamefont {Guan}, \citenamefont {Diao} \emph {et~al.}}]{ref18B}%
  \BibitemOpen
  \bibfield  {author} {\bibinfo {author} {\bibfnamefont {Y.-J.}\ \bibnamefont
  {Wang}}, \bibinfo {author} {\bibfnamefont {F.-H.}\ \bibnamefont {Guan}},
  \bibinfo {author} {\bibfnamefont {X.-Y.}\ \bibnamefont {Diao}},  \emph
  {et~al.},\ }\href {\doibase https://doi.org/10.1007/s41365-020-00842-2}
  {\bibfield  {journal} {\bibinfo  {journal} {Nucl. Sci. Tech.}\ }\textbf
  {\bibinfo {volume} {32}},\ \bibinfo {pages} {4} (\bibinfo {year}
  {2021})}\BibitemShut {NoStop}%
\bibitem [{\citenamefont {Wang}\ \emph {et~al.}(2018)\citenamefont {Wang},
  \citenamefont {Ma}, \citenamefont {Zhang},\ and\ \citenamefont
  {Zhang}}]{ref_wangtt}%
  \BibitemOpen
  \bibfield  {author} {\bibinfo {author} {\bibfnamefont {T.-T.}\ \bibnamefont
  {Wang}}, \bibinfo {author} {\bibfnamefont {Y.-G.}\ \bibnamefont {Ma}},
  \bibinfo {author} {\bibfnamefont {C.-J.}\ \bibnamefont {Zhang}}, \ and\
  \bibinfo {author} {\bibfnamefont {Z.-Q.}\ \bibnamefont {Zhang}},\ }\href
  {\doibase 10.1103/PhysRevC.97.034617} {\bibfield  {journal} {\bibinfo
  {journal} {Phys. Rev. C}\ }\textbf {\bibinfo {volume} {97}},\ \bibinfo
  {pages} {034617} (\bibinfo {year} {2018})},\ \Eprint
  {http://arxiv.org/abs/1803.08825} {arXiv:1803.08825 [nucl-th]} \BibitemShut
  {NoStop}%
\bibitem [{\citenamefont {Zhou}\ and\ \citenamefont {Fang}(2020)}]{ref_zhou}%
  \BibitemOpen
  \bibfield  {author} {\bibinfo {author} {\bibfnamefont {L.}~\bibnamefont
  {Zhou}}\ and\ \bibinfo {author} {\bibfnamefont {D.-Q.}\ \bibnamefont
  {Fang}},\ }\href {\doibase 10.1007/s41365-020-00759-w} {\bibfield  {journal}
  {\bibinfo  {journal} {Nucl. Sci. Tech.}\ }\textbf {\bibinfo {volume} {31}},\
  \bibinfo {pages} {52} (\bibinfo {year} {2020})}\BibitemShut {NoStop}%
\bibitem [{\citenamefont {Xi}\ \emph {et~al.}(2020)\citenamefont {Xi},
  \citenamefont {Zhang}, \citenamefont {Zhang},\ and\ \citenamefont
  {Ma}}]{XiBS}%
  \BibitemOpen
  \bibfield  {author} {\bibinfo {author} {\bibfnamefont {B.-S.}\ \bibnamefont
  {Xi}}, \bibinfo {author} {\bibfnamefont {Z.-Q.}\ \bibnamefont {Zhang}},
  \bibinfo {author} {\bibfnamefont {S.}~\bibnamefont {Zhang}}, \ and\ \bibinfo
  {author} {\bibfnamefont {Y.-G.}\ \bibnamefont {Ma}},\ }\href {\doibase
  https://doi.org/10.1103/PhysRevC.102.064901} {\bibfield  {journal} {\bibinfo
  {journal} {Phys. Rev. C}\ }\textbf {\bibinfo {volume} {102}},\ \bibinfo
  {pages} {064901} (\bibinfo {year} {2020})}\BibitemShut {NoStop}%
\bibitem [{\citenamefont {Lisa}\ \emph
  {et~al.}(2000{\natexlab{a}})\citenamefont {Lisa}, \citenamefont {Heinz},\
  and\ \citenamefont {Wiedemann}}]{ref59}%
  \BibitemOpen
  \bibfield  {author} {\bibinfo {author} {\bibfnamefont {M.~A.}\ \bibnamefont
  {Lisa}}, \bibinfo {author} {\bibfnamefont {U.~W.}\ \bibnamefont {Heinz}}, \
  and\ \bibinfo {author} {\bibfnamefont {U.~A.}\ \bibnamefont {Wiedemann}},\
  }\href {\doibase 10.1016/S0370-2693(00)00952-7} {\bibfield  {journal}
  {\bibinfo  {journal} {Phys. Lett. B}\ }\textbf {\bibinfo {volume} {489}},\
  \bibinfo {pages} {287} (\bibinfo {year} {2000}{\natexlab{a}})},\ \Eprint
  {http://arxiv.org/abs/nucl-th/0003022} {arXiv:nucl-th/0003022} \BibitemShut
  {NoStop}%
\bibitem [{\citenamefont {Adamczewski-Musch}\ \emph
  {et~al.}(2020{\natexlab{a}})\citenamefont {Adamczewski-Musch}, \citenamefont
  {Arnold}, \citenamefont {Behnke} \emph {et~al.}}]{HADES}%
  \BibitemOpen
  \bibfield  {author} {\bibinfo {author} {\bibfnamefont {J.}~\bibnamefont
  {Adamczewski-Musch}}, \bibinfo {author} {\bibfnamefont {O.}~\bibnamefont
  {Arnold}}, \bibinfo {author} {\bibfnamefont {C.}~\bibnamefont {Behnke}},
  \emph {et~al.} (\bibinfo {collaboration} {HADES Collaboration}),\ }\href
  {\doibase 10.1140/epja/s10050-020-00237-2} {\bibfield  {journal} {\bibinfo
  {journal} {Eur. Phys. J. A}\ }\textbf {\bibinfo {volume} {56}},\ \bibinfo
  {pages} {259} (\bibinfo {year} {2020}{\natexlab{a}})}\BibitemShut {NoStop}%
\bibitem [{\citenamefont {Pratt}\ \emph {et~al.}(1994)\citenamefont {Pratt},
  \citenamefont {Sullivan}, \citenamefont {Sorge} \emph
  {et~al.}}]{PRATT1994103}%
  \BibitemOpen
  \bibfield  {author} {\bibinfo {author} {\bibfnamefont {S.}~\bibnamefont
  {Pratt}}, \bibinfo {author} {\bibfnamefont {J.}~\bibnamefont {Sullivan}},
  \bibinfo {author} {\bibfnamefont {H.}~\bibnamefont {Sorge}},  \emph
  {et~al.},\ }\href {\doibase 10.1016/0375-9474(94)90614-9} {\bibfield
  {journal} {\bibinfo  {journal} {Nucl. Phys. A}\ }\textbf {\bibinfo {volume}
  {566}},\ \bibinfo {pages} {103 } (\bibinfo {year} {1994})}\BibitemShut
  {NoStop}%
\bibitem [{\citenamefont {Aichelin}(1991)}]{ref_aichelin}%
  \BibitemOpen
  \bibfield  {author} {\bibinfo {author} {\bibfnamefont {J.}~\bibnamefont
  {Aichelin}},\ }\href {\doibase https://doi.org/10.1016/0370-1573(91)90094-3}
  {\bibfield  {journal} {\bibinfo  {journal} {Phys. Rep.}\ }\textbf {\bibinfo
  {volume} {202}},\ \bibinfo {pages} {233} (\bibinfo {year}
  {1991})}\BibitemShut {NoStop}%
\bibitem [{\citenamefont {Ma}\ and\ \citenamefont {Shen}(1995)}]{ref_ma}%
  \BibitemOpen
  \bibfield  {author} {\bibinfo {author} {\bibfnamefont {Y.~G.}\ \bibnamefont
  {Ma}}\ and\ \bibinfo {author} {\bibfnamefont {W.~Q.}\ \bibnamefont {Shen}},\
  }\href {\doibase 10.1103/PhysRevC.51.710} {\bibfield  {journal} {\bibinfo
  {journal} {Phys. Rev. C}\ }\textbf {\bibinfo {volume} {51}},\ \bibinfo
  {pages} {710} (\bibinfo {year} {1995})}\BibitemShut {NoStop}%
\bibitem [{\citenamefont {Feng}(2018)}]{ref_feng}%
  \BibitemOpen
  \bibfield  {author} {\bibinfo {author} {\bibfnamefont {Z.-Q.}\ \bibnamefont
  {Feng}},\ }\href {\doibase 10.1007/s41365-018-0379-z} {\bibfield  {journal}
  {\bibinfo  {journal} {Nucl. Sci. Tech.}\ }\textbf {\bibinfo {volume} {29}},\
  \bibinfo {pages} {40} (\bibinfo {year} {2018})}\BibitemShut {NoStop}%
\bibitem [{\citenamefont {Zhang}\ \emph {et~al.}(2021)\citenamefont {Zhang},
  \citenamefont {Cheng},\ and\ \citenamefont {Feng}}]{ref_feng2}%
  \BibitemOpen
  \bibfield  {author} {\bibinfo {author} {\bibfnamefont {D.-C.}\ \bibnamefont
  {Zhang}}, \bibinfo {author} {\bibfnamefont {H.-G.}\ \bibnamefont {Cheng}}, \
  and\ \bibinfo {author} {\bibfnamefont {Z.-Q.}\ \bibnamefont {Feng}},\ }\href
  {\doibase 10.1088/0256-307X/38/9/092501} {\bibfield  {journal} {\bibinfo
  {journal} {Chin. Phys. Lett.}\ }\textbf {\bibinfo {volume} {38}},\ \bibinfo
  {pages} {092501} (\bibinfo {year} {2021})}\BibitemShut {NoStop}%
\bibitem [{\citenamefont {Guo}\ \emph {et~al.}(2020)\citenamefont {Guo},
  \citenamefont {Su},\ and\ \citenamefont {Zhu}}]{ref_guo}%
  \BibitemOpen
  \bibfield  {author} {\bibinfo {author} {\bibfnamefont {C.-C.}\ \bibnamefont
  {Guo}}, \bibinfo {author} {\bibfnamefont {J.}~\bibnamefont {Su}}, \ and\
  \bibinfo {author} {\bibfnamefont {L.}~\bibnamefont {Zhu}},\ }\href {\doibase
  10.1007/s41365-020-00832-4} {\bibfield  {journal} {\bibinfo  {journal} {Nucl.
  Sci. Tech.}\ }\textbf {\bibinfo {volume} {31}},\ \bibinfo {pages} {123}
  (\bibinfo {year} {2020})}\BibitemShut {NoStop}%
\bibitem [{\citenamefont {Hartnack}\ \emph {et~al.}(1998)\citenamefont
  {Hartnack}, \citenamefont {Puri}, \citenamefont {Aichelin}, \citenamefont
  {Konopka}, \citenamefont {Bass}, \citenamefont {Stoecker},\ and\
  \citenamefont {Greiner}}]{ref38}%
  \BibitemOpen
  \bibfield  {author} {\bibinfo {author} {\bibfnamefont {C.}~\bibnamefont
  {Hartnack}}, \bibinfo {author} {\bibfnamefont {R.~K.}\ \bibnamefont {Puri}},
  \bibinfo {author} {\bibfnamefont {J.}~\bibnamefont {Aichelin}}, \bibinfo
  {author} {\bibfnamefont {J.}~\bibnamefont {Konopka}}, \bibinfo {author}
  {\bibfnamefont {S.}~\bibnamefont {Bass}}, \bibinfo {author} {\bibfnamefont
  {H.}~\bibnamefont {Stoecker}}, \ and\ \bibinfo {author} {\bibfnamefont
  {W.}~\bibnamefont {Greiner}},\ }\href {\doibase 10.1007/s100500050045}
  {\bibfield  {journal} {\bibinfo  {journal} {Eur. Phys. J. A}\ }\textbf
  {\bibinfo {volume} {1}},\ \bibinfo {pages} {151} (\bibinfo {year} {1998})},\
  \Eprint {http://arxiv.org/abs/nucl-th/9811015} {arXiv:nucl-th/9811015}
  \BibitemShut {NoStop}%
\bibitem [{\citenamefont {Liu}\ \emph {et~al.}(2017)\citenamefont {Liu},
  \citenamefont {Ma}, \citenamefont {Bonasera}, \citenamefont {Deng},
  \citenamefont {Lopez},\ and\ \citenamefont
  {Veselsk\'y}}]{PhysRevC.96.064604}%
  \BibitemOpen
  \bibfield  {author} {\bibinfo {author} {\bibfnamefont {H.~L.}\ \bibnamefont
  {Liu}}, \bibinfo {author} {\bibfnamefont {Y.~G.}\ \bibnamefont {Ma}},
  \bibinfo {author} {\bibfnamefont {A.}~\bibnamefont {Bonasera}}, \bibinfo
  {author} {\bibfnamefont {X.~G.}\ \bibnamefont {Deng}}, \bibinfo {author}
  {\bibfnamefont {O.}~\bibnamefont {Lopez}}, \ and\ \bibinfo {author}
  {\bibfnamefont {M.}~\bibnamefont {Veselsk\'y}},\ }\href {\doibase
  10.1103/PhysRevC.96.064604} {\bibfield  {journal} {\bibinfo  {journal} {Phys.
  Rev. C}\ }\textbf {\bibinfo {volume} {96}},\ \bibinfo {pages} {064604}
  (\bibinfo {year} {2017})}\BibitemShut {NoStop}%
\bibitem [{\citenamefont {Yu}\ \emph {et~al.}(2020)\citenamefont {Yu},
  \citenamefont {Fang},\ and\ \citenamefont {Ma}}]{ref_yu}%
  \BibitemOpen
  \bibfield  {author} {\bibinfo {author} {\bibfnamefont {H.}~\bibnamefont
  {Yu}}, \bibinfo {author} {\bibfnamefont {D.-Q.}\ \bibnamefont {Fang}}, \ and\
  \bibinfo {author} {\bibfnamefont {Y.-G.}\ \bibnamefont {Ma}},\ }\href
  {\doibase 10.1007/s41365-020-00766-x} {\bibfield  {journal} {\bibinfo
  {journal} {Nucl. Sci. Tech.}\ }\textbf {\bibinfo {volume} {31}},\ \bibinfo
  {pages} {61} (\bibinfo {year} {2020})}\BibitemShut {NoStop}%
\bibitem [{\citenamefont {Westfall}\ \emph {et~al.}(1993)\citenamefont
  {Westfall}, \citenamefont {Bauer}, \citenamefont {Craig} \emph
  {et~al.}}]{ref_westfall}%
  \BibitemOpen
  \bibfield  {author} {\bibinfo {author} {\bibfnamefont {G.~D.}\ \bibnamefont
  {Westfall}}, \bibinfo {author} {\bibfnamefont {W.}~\bibnamefont {Bauer}},
  \bibinfo {author} {\bibfnamefont {D.}~\bibnamefont {Craig}},  \emph
  {et~al.},\ }\href {\doibase https://doi.org/10.1103/PhysRevLett.71.1986}
  {\bibfield  {journal} {\bibinfo  {journal} {Phys. Rev. Lett.}\ }\textbf
  {\bibinfo {volume} {71}},\ \bibinfo {pages} {1986} (\bibinfo {year}
  {1993})}\BibitemShut {NoStop}%
\bibitem [{\citenamefont {Chen}\ \emph {et~al.}(1968)\citenamefont {Chen},
  \citenamefont {Fraenkel}, \citenamefont {Friedlander}, \citenamefont
  {Grover}, \citenamefont {Miller},\ and\ \citenamefont {Shimamoto}}]{ref40}%
  \BibitemOpen
  \bibfield  {author} {\bibinfo {author} {\bibfnamefont {K.}~\bibnamefont
  {Chen}}, \bibinfo {author} {\bibfnamefont {Z.}~\bibnamefont {Fraenkel}},
  \bibinfo {author} {\bibfnamefont {G.}~\bibnamefont {Friedlander}}, \bibinfo
  {author} {\bibfnamefont {J.~R.}\ \bibnamefont {Grover}}, \bibinfo {author}
  {\bibfnamefont {J.~M.}\ \bibnamefont {Miller}}, \ and\ \bibinfo {author}
  {\bibfnamefont {Y.}~\bibnamefont {Shimamoto}},\ }\href {\doibase
  10.1103/PhysRev.166.949} {\bibfield  {journal} {\bibinfo  {journal} {Phys.
  Rev.}\ }\textbf {\bibinfo {volume} {166}},\ \bibinfo {pages} {949} (\bibinfo
  {year} {1968})}\BibitemShut {NoStop}%
\bibitem [{\citenamefont {Zhang}\ and\ \citenamefont {Su}(2020)}]{ref_zhang}%
  \BibitemOpen
  \bibfield  {author} {\bibinfo {author} {\bibfnamefont {F.}~\bibnamefont
  {Zhang}}\ and\ \bibinfo {author} {\bibfnamefont {J.}~\bibnamefont {Su}},\
  }\href {\doibase 10.1007/s41365-020-00787-6} {\bibfield  {journal} {\bibinfo
  {journal} {Nucl. Sci. Tech.}\ }\textbf {\bibinfo {volume} {31}},\ \bibinfo
  {pages} {77} (\bibinfo {year} {2020})}\BibitemShut {NoStop}%
\bibitem [{\citenamefont {Miller}\ \emph {et~al.}(2007)\citenamefont {Miller},
  \citenamefont {Reygers}, \citenamefont {Sanders},\ and\ \citenamefont
  {Steinberg}}]{ref42}%
  \BibitemOpen
  \bibfield  {author} {\bibinfo {author} {\bibfnamefont {M.~L.}\ \bibnamefont
  {Miller}}, \bibinfo {author} {\bibfnamefont {K.}~\bibnamefont {Reygers}},
  \bibinfo {author} {\bibfnamefont {S.~J.}\ \bibnamefont {Sanders}}, \ and\
  \bibinfo {author} {\bibfnamefont {P.}~\bibnamefont {Steinberg}},\ }\href
  {\doibase 10.1146/annurev.nucl.57.090506.123020} {\bibfield  {journal}
  {\bibinfo  {journal} {Ann. Rev. Nucl. Part. Sci.}\ }\textbf {\bibinfo
  {volume} {57}},\ \bibinfo {pages} {205} (\bibinfo {year} {2007})},\ \Eprint
  {http://arxiv.org/abs/nucl-ex/0701025} {arXiv:nucl-ex/0701025} \BibitemShut
  {NoStop}%
\bibitem [{\citenamefont {Lacey}\ \emph {et~al.}(2011)\citenamefont {Lacey},
  \citenamefont {Wei}, \citenamefont {Ajitanand},\ and\ \citenamefont
  {Taranenko}}]{ref43}%
  \BibitemOpen
  \bibfield  {author} {\bibinfo {author} {\bibfnamefont {R.~A.}\ \bibnamefont
  {Lacey}}, \bibinfo {author} {\bibfnamefont {R.}~\bibnamefont {Wei}}, \bibinfo
  {author} {\bibfnamefont {N.}~\bibnamefont {Ajitanand}}, \ and\ \bibinfo
  {author} {\bibfnamefont {A.}~\bibnamefont {Taranenko}},\ }\href {\doibase
  10.1103/PhysRevC.83.044902} {\bibfield  {journal} {\bibinfo  {journal} {Phys.
  Rev. C}\ }\textbf {\bibinfo {volume} {83}},\ \bibinfo {pages} {044902}
  (\bibinfo {year} {2011})},\ \Eprint {http://arxiv.org/abs/1009.5230}
  {arXiv:1009.5230 [nucl-ex]} \BibitemShut {NoStop}%
\bibitem [{\citenamefont {Adare}\ \emph {et~al.}(2014)\citenamefont {Adare}
  \emph {et~al.}}]{ref44}%
  \BibitemOpen
  \bibfield  {author} {\bibinfo {author} {\bibfnamefont {A.}~\bibnamefont
  {Adare}} \emph {et~al.} (\bibinfo {collaboration} {PHENIX Collaboration}),\
  }\href {\doibase 10.1103/PhysRevC.90.034902} {\bibfield  {journal} {\bibinfo
  {journal} {Phys. Rev. C}\ }\textbf {\bibinfo {volume} {90}},\ \bibinfo
  {pages} {034902} (\bibinfo {year} {2014})},\ \Eprint
  {http://arxiv.org/abs/1310.4793} {arXiv:1310.4793 [nucl-ex]} \BibitemShut
  {NoStop}%
\bibitem [{\citenamefont {Pratt}(1984)}]{ref45}%
  \BibitemOpen
  \bibfield  {author} {\bibinfo {author} {\bibfnamefont {S.}~\bibnamefont
  {Pratt}},\ }\href {\doibase 10.1103/PhysRevLett.53.1219} {\bibfield
  {journal} {\bibinfo  {journal} {Phys. Rev. Lett.}\ }\textbf {\bibinfo
  {volume} {53}},\ \bibinfo {pages} {1219} (\bibinfo {year}
  {1984})}\BibitemShut {NoStop}%
\bibitem [{\citenamefont {Bowler}(1991)}]{ref46}%
  \BibitemOpen
  \bibfield  {author} {\bibinfo {author} {\bibfnamefont {M.~G.}\ \bibnamefont
  {Bowler}},\ }\href {\doibase 10.1016/0370-2693(91)91541-3} {\bibfield
  {journal} {\bibinfo  {journal} {Phys. Lett. B}\ }\textbf {\bibinfo {volume}
  {270}},\ \bibinfo {pages} {69} (\bibinfo {year} {1991})}\BibitemShut
  {NoStop}%
\bibitem [{\citenamefont {Sinyukov}\ \emph {et~al.}(1998)\citenamefont
  {Sinyukov}, \citenamefont {Lednicky}, \citenamefont {Akkelin}, \citenamefont
  {Pluta},\ and\ \citenamefont {Erazmus}}]{ref47}%
  \BibitemOpen
  \bibfield  {author} {\bibinfo {author} {\bibfnamefont {Y.}~\bibnamefont
  {Sinyukov}}, \bibinfo {author} {\bibfnamefont {R.}~\bibnamefont {Lednicky}},
  \bibinfo {author} {\bibfnamefont {S.~V.}\ \bibnamefont {Akkelin}}, \bibinfo
  {author} {\bibfnamefont {J.}~\bibnamefont {Pluta}}, \ and\ \bibinfo {author}
  {\bibfnamefont {B.}~\bibnamefont {Erazmus}},\ }\href {\doibase
  10.1016/S0370-2693(98)00653-4} {\bibfield  {journal} {\bibinfo  {journal}
  {Phys. Lett. B}\ }\textbf {\bibinfo {volume} {432}},\ \bibinfo {pages} {248}
  (\bibinfo {year} {1998})}\BibitemShut {NoStop}%
\bibitem [{\citenamefont {Ajitanand}\ \emph {et~al.}(2014)\citenamefont
  {Ajitanand} \emph {et~al.}}]{ref48}%
  \BibitemOpen
  \bibfield  {author} {\bibinfo {author} {\bibfnamefont {N.~N.}\ \bibnamefont
  {Ajitanand}} \emph {et~al.} (\bibinfo {collaboration} {PHENIX
  Collaboration}),\ }\href {\doibase 10.1016/j.nuclphysa.2014.08.054}
  {\bibfield  {journal} {\bibinfo  {journal} {Nucl. Phys. A}\ }\textbf
  {\bibinfo {volume} {931}},\ \bibinfo {pages} {1082} (\bibinfo {year}
  {2014})},\ \Eprint {http://arxiv.org/abs/1404.5291} {arXiv:1404.5291
  [nucl-ex]} \BibitemShut {NoStop}%
\bibitem [{\citenamefont {Podgoretsky}(1983)}]{ref49}%
  \BibitemOpen
  \bibfield  {author} {\bibinfo {author} {\bibfnamefont {M.~I.}\ \bibnamefont
  {Podgoretsky}},\ }\href@noop {} {\bibfield  {journal} {\bibinfo  {journal}
  {Sov. J. Nucl. Phys.}\ }\textbf {\bibinfo {volume} {37}},\ \bibinfo {pages}
  {272} (\bibinfo {year} {1983})}\BibitemShut {NoStop}%
\bibitem [{\citenamefont {Pratt}(1986)}]{ref50}%
  \BibitemOpen
  \bibfield  {author} {\bibinfo {author} {\bibfnamefont {S.}~\bibnamefont
  {Pratt}},\ }\href {\doibase 10.1103/PhysRevD.33.1314} {\bibfield  {journal}
  {\bibinfo  {journal} {Phys. Rev. D}\ }\textbf {\bibinfo {volume} {33}},\
  \bibinfo {pages} {1314} (\bibinfo {year} {1986})}\BibitemShut {NoStop}%
\bibitem [{\citenamefont {Yu}\ \emph {et~al.}(2008)\citenamefont {Yu} \emph
  {et~al.}}]{ref51}%
  \BibitemOpen
  \bibfield  {author} {\bibinfo {author} {\bibfnamefont {L.-L.}\ \bibnamefont
  {Yu}} \emph {et~al.},\ }\href {\doibase 10.1088/1674-1137/32/11/010}
  {\bibfield  {journal} {\bibinfo  {journal} {Chin. Phys. C}\ }\textbf
  {\bibinfo {volume} {32}},\ \bibinfo {pages} {897} (\bibinfo {year}
  {2008})}\BibitemShut {NoStop}%
\bibitem [{\citenamefont {Zhang}(2011)}]{ref52}%
  \BibitemOpen
  \bibfield  {author} {\bibinfo {author} {\bibfnamefont {W.-N.}\ \bibnamefont
  {Zhang}},\ }\href {\doibase 10.1134/S1547477111090408} {\bibfield  {journal}
  {\bibinfo  {journal} {Phys. Part. Nucl. Lett.}\ }\textbf {\bibinfo {volume}
  {8}},\ \bibinfo {pages} {977} (\bibinfo {year} {2011})},\ \Eprint
  {http://arxiv.org/abs/1012.5558} {arXiv:1012.5558 [nucl-th]} \BibitemShut
  {NoStop}%
\bibitem [{\citenamefont {Zhang}\ \emph {et~al.}(2004)\citenamefont {Zhang},
  \citenamefont {Efaaf},\ and\ \citenamefont {Wong}}]{ref53}%
  \BibitemOpen
  \bibfield  {author} {\bibinfo {author} {\bibfnamefont {W.~N.}\ \bibnamefont
  {Zhang}}, \bibinfo {author} {\bibfnamefont {M.~J.}\ \bibnamefont {Efaaf}}, \
  and\ \bibinfo {author} {\bibfnamefont {C.-Y.}\ \bibnamefont {Wong}},\ }\href
  {\doibase 10.1103/PhysRevC.70.024903} {\bibfield  {journal} {\bibinfo
  {journal} {Phys. Rev. C}\ }\textbf {\bibinfo {volume} {70}},\ \bibinfo
  {pages} {024903} (\bibinfo {year} {2004})},\ \Eprint
  {http://arxiv.org/abs/hep-ph/0409013} {arXiv:hep-ph/0409013} \BibitemShut
  {NoStop}%
\bibitem [{\citenamefont {Adamczyk}\ \emph
  {et~al.}(2015{\natexlab{b}})\citenamefont {Adamczyk}, \citenamefont {Adkins},
  \citenamefont {Agakishiev} \emph {et~al.}}]{ref54}%
  \BibitemOpen
  \bibfield  {author} {\bibinfo {author} {\bibfnamefont {L.~L.}\ \bibnamefont
  {Adamczyk}}, \bibinfo {author} {\bibfnamefont {J.~K.}\ \bibnamefont
  {Adkins}}, \bibinfo {author} {\bibfnamefont {G.}~\bibnamefont {Agakishiev}},
  \emph {et~al.} (\bibinfo {collaboration} {STAR Collaboration}),\ }\href
  {https://doi.org/10.1103/PhysRevC.88.014904} {\bibfield  {journal} {\bibinfo
  {journal} {Phys. Rev. C}\ }\textbf {\bibinfo {volume} {92}},\ \bibinfo
  {pages} {014904} (\bibinfo {year} {2015}{\natexlab{b}})},\ \Eprint
  {http://arxiv.org/abs/1403.4972} {arXiv:1403.4972 [nucl-ex]} \BibitemShut
  {NoStop}%
\bibitem [{\citenamefont {Retiere}\ and\ \citenamefont
  {Lisa}(2004{\natexlab{a}})}]{ref55}%
  \BibitemOpen
  \bibfield  {author} {\bibinfo {author} {\bibfnamefont {F.}~\bibnamefont
  {Retiere}}\ and\ \bibinfo {author} {\bibfnamefont {M.~A.}\ \bibnamefont
  {Lisa}},\ }\href {\doibase 10.1103/PhysRevC.70.044907} {\bibfield  {journal}
  {\bibinfo  {journal} {Phys. Rev. C}\ }\textbf {\bibinfo {volume} {70}},\
  \bibinfo {pages} {044907} (\bibinfo {year} {2004}{\natexlab{a}})},\ \Eprint
  {http://arxiv.org/abs/nucl-th/0312024} {arXiv:nucl-th/0312024} \BibitemShut
  {NoStop}%
\bibitem [{\citenamefont {Adamczewski-Musch}\ \emph
  {et~al.}(2020{\natexlab{b}})\citenamefont {Adamczewski-Musch} \emph
  {et~al.}}]{ref56}%
  \BibitemOpen
  \bibfield  {author} {\bibinfo {author} {\bibfnamefont {J.}~\bibnamefont
  {Adamczewski-Musch}} \emph {et~al.} (\bibinfo {collaboration} {HADES
  Collaboration}),\ }\href {\doibase 10.1140/epja/s10050-020-00116-w}
  {\bibfield  {journal} {\bibinfo  {journal} {Eur. Phys. J. A}\ }\textbf
  {\bibinfo {volume} {56}},\ \bibinfo {pages} {140} (\bibinfo {year}
  {2020}{\natexlab{b}})},\ \Eprint {http://arxiv.org/abs/1910.07885}
  {arXiv:1910.07885 [nucl-ex]} \BibitemShut {NoStop}%
\bibitem [{\citenamefont {Wiedemann}\ and\ \citenamefont
  {Heinz}(1999)}]{ref57}%
  \BibitemOpen
  \bibfield  {author} {\bibinfo {author} {\bibfnamefont {U.~A.}\ \bibnamefont
  {Wiedemann}}\ and\ \bibinfo {author} {\bibfnamefont {U.~W.}\ \bibnamefont
  {Heinz}},\ }\href {\doibase 10.1016/S0370-1573(99)00032-0} {\bibfield
  {journal} {\bibinfo  {journal} {Phys. Rept.}\ }\textbf {\bibinfo {volume}
  {319}},\ \bibinfo {pages} {145} (\bibinfo {year} {1999})},\ \Eprint
  {http://arxiv.org/abs/nucl-th/9901094} {arXiv:nucl-th/9901094} \BibitemShut
  {NoStop}%
\bibitem [{\citenamefont {Lisa}\ \emph
  {et~al.}(2000{\natexlab{b}})\citenamefont {Lisa} \emph {et~al.}}]{ref58}%
  \BibitemOpen
  \bibfield  {author} {\bibinfo {author} {\bibfnamefont {M.}~\bibnamefont
  {Lisa}} \emph {et~al.} (\bibinfo {collaboration} {E895}),\ }\href {\doibase
  10.1016/S0370-2693(00)01280-6} {\bibfield  {journal} {\bibinfo  {journal}
  {Phys. Lett. B}\ }\textbf {\bibinfo {volume} {496}},\ \bibinfo {pages} {1}
  (\bibinfo {year} {2000}{\natexlab{b}})},\ \Eprint
  {http://arxiv.org/abs/nucl-ex/0007022} {arXiv:nucl-ex/0007022} \BibitemShut
  {NoStop}%
\bibitem [{\citenamefont {Voloshin}\ and\ \citenamefont
  {Cleland}(1996)}]{ref60}%
  \BibitemOpen
  \bibfield  {author} {\bibinfo {author} {\bibfnamefont {S.~A.}\ \bibnamefont
  {Voloshin}}\ and\ \bibinfo {author} {\bibfnamefont {W.~E.}\ \bibnamefont
  {Cleland}},\ }\href {\doibase 10.1103/PhysRevC.53.896} {\bibfield  {journal}
  {\bibinfo  {journal} {Phys. Rev. C}\ }\textbf {\bibinfo {volume} {53}},\
  \bibinfo {pages} {896} (\bibinfo {year} {1996})},\ \Eprint
  {http://arxiv.org/abs/nucl-th/9509025} {arXiv:nucl-th/9509025} \BibitemShut
  {NoStop}%
\bibitem [{\citenamefont {Heiselberg}(1999)}]{ref61}%
  \BibitemOpen
  \bibfield  {author} {\bibinfo {author} {\bibfnamefont {H.}~\bibnamefont
  {Heiselberg}},\ }\href {\doibase 10.1103/PhysRevLett.82.2052} {\bibfield
  {journal} {\bibinfo  {journal} {Phys. Rev. Lett.}\ }\textbf {\bibinfo
  {volume} {82}},\ \bibinfo {pages} {2052} (\bibinfo {year} {1999})},\ \Eprint
  {http://arxiv.org/abs/nucl-th/9809077} {arXiv:nucl-th/9809077} \BibitemShut
  {NoStop}%
\bibitem [{\citenamefont {Heiselberg}\ and\ \citenamefont
  {Levy}(1999)}]{ref62}%
  \BibitemOpen
  \bibfield  {author} {\bibinfo {author} {\bibfnamefont {H.}~\bibnamefont
  {Heiselberg}}\ and\ \bibinfo {author} {\bibfnamefont {A.-M.}\ \bibnamefont
  {Levy}},\ }\href {\doibase 10.1103/PhysRevC.59.2716} {\bibfield  {journal}
  {\bibinfo  {journal} {Phys. Rev. C}\ }\textbf {\bibinfo {volume} {59}},\
  \bibinfo {pages} {2716} (\bibinfo {year} {1999})},\ \Eprint
  {http://arxiv.org/abs/nucl-th/9812034} {arXiv:nucl-th/9812034} \BibitemShut
  {NoStop}%
\bibitem [{\citenamefont {Li}\ \emph {et~al.}(2008)\citenamefont {Li},
  \citenamefont {Chen},\ and\ \citenamefont {Ko}}]{BALi}%
  \BibitemOpen
  \bibfield  {author} {\bibinfo {author} {\bibfnamefont {B.-A.}\ \bibnamefont
  {Li}}, \bibinfo {author} {\bibfnamefont {L.-W.}\ \bibnamefont {Chen}}, \ and\
  \bibinfo {author} {\bibfnamefont {C.~M.}\ \bibnamefont {Ko}},\ }\href
  {\doibase 0.1016/j.physrep.2008.04.005} {\bibfield  {journal} {\bibinfo
  {journal} {Physics Reports}\ }\textbf {\bibinfo {volume} {464}},\ \bibinfo
  {pages} {113} (\bibinfo {year} {2008})}\BibitemShut {NoStop}%
\bibitem [{\citenamefont {Li}\ and\ \citenamefont {Bauer}(1991)}]{BALi2}%
  \BibitemOpen
  \bibfield  {author} {\bibinfo {author} {\bibfnamefont {B.~A.}\ \bibnamefont
  {Li}}\ and\ \bibinfo {author} {\bibfnamefont {W.}~\bibnamefont {Bauer}},\
  }\href {\doibase 10.1016/0370-2693(91)91165-R} {\bibfield  {journal}
  {\bibinfo  {journal} {Phys. Lett. B}\ }\textbf {\bibinfo {volume} {254}},\
  \bibinfo {pages} {335} (\bibinfo {year} {1991})}\BibitemShut {NoStop}%
\bibitem [{\citenamefont {Wang}\ \emph {et~al.}(2020)\citenamefont {Wang},
  \citenamefont {Zhang}, \citenamefont {Chen}, \citenamefont {Ko},\ and\
  \citenamefont {Ma}}]{WangRui}%
  \BibitemOpen
  \bibfield  {author} {\bibinfo {author} {\bibfnamefont {R.}~\bibnamefont
  {Wang}}, \bibinfo {author} {\bibfnamefont {Z.}~\bibnamefont {Zhang}},
  \bibinfo {author} {\bibfnamefont {L.-W.}\ \bibnamefont {Chen}}, \bibinfo
  {author} {\bibfnamefont {C.~M.}\ \bibnamefont {Ko}}, \ and\ \bibinfo {author}
  {\bibfnamefont {Y.-G.}\ \bibnamefont {Ma}},\ }\href {\doibase
  https://doi.org/10.1016/j.physletb.2020.135532} {\bibfield  {journal}
  {\bibinfo  {journal} {Phys. Lett. B}\ }\textbf {\bibinfo {volume} {807}},\
  \bibinfo {pages} {135532} (\bibinfo {year} {2020})}\BibitemShut {NoStop}%
\bibitem [{\citenamefont {Godbey}\ \emph {et~al.}(2021)\citenamefont {Godbey},
  \citenamefont {Zhang}, \citenamefont {Holt},\ and\ \citenamefont
  {Ko}}]{2107.13384}%
  \BibitemOpen
  \bibfield  {author} {\bibinfo {author} {\bibfnamefont {K.}~\bibnamefont
  {Godbey}}, \bibinfo {author} {\bibfnamefont {Z.}~\bibnamefont {Zhang}},
  \bibinfo {author} {\bibfnamefont {J.~W.}\ \bibnamefont {Holt}}, \ and\
  \bibinfo {author} {\bibfnamefont {C.~M.}\ \bibnamefont {Ko}},\ }\href
  {\doibase 10.48550/arxiv.2107.13384} {\  (\bibinfo {year} {2021}),\
  10.48550/arxiv.2107.13384}\BibitemShut {NoStop}%
\bibitem [{\citenamefont {Lv}\ \emph {et~al.}(2017)\citenamefont {Lv},
  \citenamefont {Ma}, \citenamefont {Chen}, \citenamefont {Fang},\ and\
  \citenamefont {Zhang}}]{Lv1}%
  \BibitemOpen
  \bibfield  {author} {\bibinfo {author} {\bibfnamefont {M.}~\bibnamefont
  {Lv}}, \bibinfo {author} {\bibfnamefont {Y.~G.}\ \bibnamefont {Ma}}, \bibinfo
  {author} {\bibfnamefont {J.~H.}\ \bibnamefont {Chen}}, \bibinfo {author}
  {\bibfnamefont {D.~Q.}\ \bibnamefont {Fang}}, \ and\ \bibinfo {author}
  {\bibfnamefont {G.~Q.}\ \bibnamefont {Zhang}},\ }\href {\doibase
  10.1103/PhysRevC.95.024614} {\bibfield  {journal} {\bibinfo  {journal} {Phys.
  Rev. C}\ }\textbf {\bibinfo {volume} {95}},\ \bibinfo {pages} {024614}
  (\bibinfo {year} {2017})}\BibitemShut {NoStop}%
\bibitem [{\citenamefont {Stock}(1986)}]{Stock}%
  \BibitemOpen
  \bibfield  {author} {\bibinfo {author} {\bibfnamefont {R.}~\bibnamefont
  {Stock}},\ }\href {\doibase 10.1016/0370-1573(86)90134-1} {\bibfield
  {journal} {\bibinfo  {journal} {Phys. Rep.}\ }\textbf {\bibinfo {volume}
  {135}},\ \bibinfo {pages} {259} (\bibinfo {year} {1986})}\BibitemShut
  {NoStop}%
\bibitem [{\citenamefont {Harabasz}\ \emph {et~al.}(2020)\citenamefont
  {Harabasz}, \citenamefont {Florkowski}, \citenamefont {Galatyuk},
  \citenamefont {Gumberidze}, \citenamefont {Ryblewski}, \citenamefont
  {Salabura},\ and\ \citenamefont {Stroth}}]{harabasz2020statistical}%
  \BibitemOpen
  \bibfield  {author} {\bibinfo {author} {\bibfnamefont {S.}~\bibnamefont
  {Harabasz}}, \bibinfo {author} {\bibfnamefont {W.}~\bibnamefont
  {Florkowski}}, \bibinfo {author} {\bibfnamefont {T.}~\bibnamefont
  {Galatyuk}}, \bibinfo {author} {\bibfnamefont {M.}~\bibnamefont
  {Gumberidze}}, \bibinfo {author} {\bibfnamefont {R.}~\bibnamefont
  {Ryblewski}}, \bibinfo {author} {\bibfnamefont {P.}~\bibnamefont {Salabura}},
  \ and\ \bibinfo {author} {\bibfnamefont {J.}~\bibnamefont {Stroth}},\ }\href
  {\doibase https://doi.org/10.1103/PhysRevC.102.054903} {\bibfield  {journal}
  {\bibinfo  {journal} {Phys. Rev. C}\ }\textbf {\bibinfo {volume} {102}},\
  \bibinfo {pages} {054903} (\bibinfo {year} {2020})}\BibitemShut {NoStop}%
\bibitem [{\citenamefont {Hong}\ \emph {et~al.}(1997)\citenamefont {Hong} \emph
  {et~al.}}]{Hong:1997ka}%
  \BibitemOpen
  \bibfield  {author} {\bibinfo {author} {\bibfnamefont {B.}~\bibnamefont
  {Hong}} \emph {et~al.} (\bibinfo {collaboration} {FOPI Collaboration}),\
  }\href {\doibase 10.1016/S0370-2693(97)00707-7} {\bibfield  {journal}
  {\bibinfo  {journal} {Phys. Lett. B}\ }\textbf {\bibinfo {volume} {407}},\
  \bibinfo {pages} {115} (\bibinfo {year} {1997})},\ \Eprint
  {http://arxiv.org/abs/nucl-ex/9706001} {arXiv:nucl-ex/9706001} \BibitemShut
  {NoStop}%
\bibitem [{\citenamefont {Barz}\ \emph {et~al.}(1998)\citenamefont {Barz},
  \citenamefont {Bondorf}, \citenamefont {Gaardh\o{}je},\ and\ \citenamefont
  {Heiselberg}}]{PhysRevC.57.2536}%
  \BibitemOpen
  \bibfield  {author} {\bibinfo {author} {\bibfnamefont {H.~W.}\ \bibnamefont
  {Barz}}, \bibinfo {author} {\bibfnamefont {J.~P.}\ \bibnamefont {Bondorf}},
  \bibinfo {author} {\bibfnamefont {J.~J.}\ \bibnamefont {Gaardh\o{}je}}, \
  and\ \bibinfo {author} {\bibfnamefont {H.}~\bibnamefont {Heiselberg}},\
  }\href {\doibase 10.1103/PhysRevC.57.2536} {\bibfield  {journal} {\bibinfo
  {journal} {Phys. Rev. C}\ }\textbf {\bibinfo {volume} {57}},\ \bibinfo
  {pages} {2536} (\bibinfo {year} {1998})}\BibitemShut {NoStop}%
\bibitem [{\citenamefont {Lv}\ \emph {et~al.}(2014)\citenamefont {Lv},
  \citenamefont {Ma}, \citenamefont {Zhang}, \citenamefont {Chen},\ and\
  \citenamefont {Fang}}]{Lv2}%
  \BibitemOpen
  \bibfield  {author} {\bibinfo {author} {\bibfnamefont {M.}~\bibnamefont
  {Lv}}, \bibinfo {author} {\bibfnamefont {Y.~G.}\ \bibnamefont {Ma}}, \bibinfo
  {author} {\bibfnamefont {G.~Q.}\ \bibnamefont {Zhang}}, \bibinfo {author}
  {\bibfnamefont {J.~H.}\ \bibnamefont {Chen}}, \ and\ \bibinfo {author}
  {\bibfnamefont {D.~Q.}\ \bibnamefont {Fang}},\ }\href {\doibase
  10.1016/j.physletb.2014.04.025} {\bibfield  {journal} {\bibinfo  {journal}
  {Phys. Lett. B}\ }\textbf {\bibinfo {volume} {733}},\ \bibinfo {pages} {105}
  (\bibinfo {year} {2014})}\BibitemShut {NoStop}%
\bibitem [{\citenamefont {Poskanzer}\ and\ \citenamefont
  {Voloshin}(1998)}]{PhysRevC.58.1671}%
  \BibitemOpen
  \bibfield  {author} {\bibinfo {author} {\bibfnamefont {A.~M.}\ \bibnamefont
  {Poskanzer}}\ and\ \bibinfo {author} {\bibfnamefont {S.~A.}\ \bibnamefont
  {Voloshin}},\ }\href {\doibase 10.1103/PhysRevC.58.1671} {\bibfield
  {journal} {\bibinfo  {journal} {Phys. Rev. C}\ }\textbf {\bibinfo {volume}
  {58}},\ \bibinfo {pages} {1671} (\bibinfo {year} {1998})}\BibitemShut
  {NoStop}%
\bibitem [{\citenamefont {Adamczyk}\ \emph {et~al.}(2013)\citenamefont
  {Adamczyk} \emph {et~al.}}]{PhysRevC.88.014904}%
  \BibitemOpen
  \bibfield  {author} {\bibinfo {author} {\bibfnamefont {L.}~\bibnamefont
  {Adamczyk}} \emph {et~al.} (\bibinfo {collaboration} {STAR Collaboration}),\
  }\href {https://link.aps.org/doi/10.1103/PhysRevC.88.014904} {\bibfield
  {journal} {\bibinfo  {journal} {Phys. Rev. C}\ }\textbf {\bibinfo {volume}
  {88}},\ \bibinfo {pages} {014904} (\bibinfo {year} {2013})}\BibitemShut
  {NoStop}%
\bibitem [{\citenamefont {Retiere}\ and\ \citenamefont
  {Lisa}(2004{\natexlab{b}})}]{retiere2004observable}%
  \BibitemOpen
  \bibfield  {author} {\bibinfo {author} {\bibfnamefont {F.}~\bibnamefont
  {Retiere}}\ and\ \bibinfo {author} {\bibfnamefont {M.~A.}\ \bibnamefont
  {Lisa}},\ }\href {\doibase 10.1103/PhysRevC.70.044907} {\bibfield  {journal}
  {\bibinfo  {journal} {Phys. Rev. C}\ }\textbf {\bibinfo {volume} {70}},\
  \bibinfo {pages} {044907} (\bibinfo {year} {2004}{\natexlab{b}})}\BibitemShut
  {NoStop}%
\bibitem [{\citenamefont {Barz}(1999)}]{barz1999combined}%
  \BibitemOpen
  \bibfield  {author} {\bibinfo {author} {\bibfnamefont {H.}~\bibnamefont
  {Barz}},\ }\href {\doibase https://doi.org/10.1103/PhysRevC.59.2214}
  {\bibfield  {journal} {\bibinfo  {journal} {Phys. Rev. C}\ }\textbf {\bibinfo
  {volume} {59}},\ \bibinfo {pages} {2214} (\bibinfo {year}
  {1999})}\BibitemShut {NoStop}%
\bibitem [{\citenamefont {Hung}\ and\ \citenamefont {Shuryak}(1995)}]{ref72}%
  \BibitemOpen
  \bibfield  {author} {\bibinfo {author} {\bibfnamefont {C.~M.}\ \bibnamefont
  {Hung}}\ and\ \bibinfo {author} {\bibfnamefont {E.~V.}\ \bibnamefont
  {Shuryak}},\ }\href {\doibase 10.1103/PhysRevLett.75.4003} {\bibfield
  {journal} {\bibinfo  {journal} {Phys. Rev. Lett.}\ }\textbf {\bibinfo
  {volume} {75}},\ \bibinfo {pages} {4003} (\bibinfo {year} {1995})},\ \Eprint
  {http://arxiv.org/abs/hep-ph/9412360} {arXiv:hep-ph/9412360} \BibitemShut
  {NoStop}%
\bibitem [{\citenamefont {Rischke}\ and\ \citenamefont
  {Gyulassy}(1996)}]{ref73}%
  \BibitemOpen
  \bibfield  {author} {\bibinfo {author} {\bibfnamefont {D.~H.}\ \bibnamefont
  {Rischke}}\ and\ \bibinfo {author} {\bibfnamefont {M.}~\bibnamefont
  {Gyulassy}},\ }\href {\doibase 10.1016/0375-9474(96)00259-X} {\bibfield
  {journal} {\bibinfo  {journal} {Nucl. Phys. A}\ }\textbf {\bibinfo {volume}
  {608}},\ \bibinfo {pages} {479} (\bibinfo {year} {1996})},\ \Eprint
  {http://arxiv.org/abs/nucl-th/9606039} {arXiv:nucl-th/9606039} \BibitemShut
  {NoStop}%
\bibitem [{\citenamefont {Chapman}\ \emph {et~al.}(1995)\citenamefont
  {Chapman}, \citenamefont {Scotto},\ and\ \citenamefont {Heinz}}]{ref75}%
  \BibitemOpen
  \bibfield  {author} {\bibinfo {author} {\bibfnamefont {S.}~\bibnamefont
  {Chapman}}, \bibinfo {author} {\bibfnamefont {P.}~\bibnamefont {Scotto}}, \
  and\ \bibinfo {author} {\bibfnamefont {U.~W.}\ \bibnamefont {Heinz}},\ }\href
  {\doibase 10.1103/PhysRevLett.74.4400} {\bibfield  {journal} {\bibinfo
  {journal} {Phys. Rev. Lett.}\ }\textbf {\bibinfo {volume} {74}},\ \bibinfo
  {pages} {4400} (\bibinfo {year} {1995})},\ \Eprint
  {http://arxiv.org/abs/hep-ph/9408207} {arXiv:hep-ph/9408207} \BibitemShut
  {NoStop}%
\bibitem [{\citenamefont {Wiedemann}\ \emph {et~al.}(1996)\citenamefont
  {Wiedemann}, \citenamefont {Scotto},\ and\ \citenamefont {Heinz}}]{ref76}%
  \BibitemOpen
  \bibfield  {author} {\bibinfo {author} {\bibfnamefont {U.~A.}\ \bibnamefont
  {Wiedemann}}, \bibinfo {author} {\bibfnamefont {P.}~\bibnamefont {Scotto}}, \
  and\ \bibinfo {author} {\bibfnamefont {U.~W.}\ \bibnamefont {Heinz}},\ }\href
  {\doibase 10.1103/PhysRevC.53.918} {\bibfield  {journal} {\bibinfo  {journal}
  {Phys. Rev. C}\ }\textbf {\bibinfo {volume} {53}},\ \bibinfo {pages} {918}
  (\bibinfo {year} {1996})},\ \Eprint {http://arxiv.org/abs/nucl-th/9508040}
  {arXiv:nucl-th/9508040} \BibitemShut {NoStop}%
\bibitem [{\citenamefont {Lacey}\ \emph {et~al.}(2013)\citenamefont {Lacey},
  \citenamefont {Gu}, \citenamefont {Gong}, \citenamefont {Reynolds},
  \citenamefont {Ajitanand}, \citenamefont {Alexander}, \citenamefont {Mwai},\
  and\ \citenamefont {Taranenko}}]{ref77}%
  \BibitemOpen
  \bibfield  {author} {\bibinfo {author} {\bibfnamefont {R.~A.}\ \bibnamefont
  {Lacey}}, \bibinfo {author} {\bibfnamefont {Y.}~\bibnamefont {Gu}}, \bibinfo
  {author} {\bibfnamefont {X.}~\bibnamefont {Gong}}, \bibinfo {author}
  {\bibfnamefont {D.}~\bibnamefont {Reynolds}}, \bibinfo {author}
  {\bibfnamefont {N.}~\bibnamefont {Ajitanand}}, \bibinfo {author}
  {\bibfnamefont {J.}~\bibnamefont {Alexander}}, \bibinfo {author}
  {\bibfnamefont {A.}~\bibnamefont {Mwai}}, \ and\ \bibinfo {author}
  {\bibfnamefont {A.}~\bibnamefont {Taranenko}},\ }\href@noop {} {\  (\bibinfo
  {year} {2013})},\ \Eprint {http://arxiv.org/abs/1301.0165} {arXiv:1301.0165
  [nucl-ex]} \BibitemShut {NoStop}%
\bibitem [{\citenamefont {Shuryak}\ and\ \citenamefont {Zahed}(2013)}]{ref78}%
  \BibitemOpen
  \bibfield  {author} {\bibinfo {author} {\bibfnamefont {E.}~\bibnamefont
  {Shuryak}}\ and\ \bibinfo {author} {\bibfnamefont {I.}~\bibnamefont
  {Zahed}},\ }\href {\doibase 10.1103/PhysRevC.88.044915} {\bibfield  {journal}
  {\bibinfo  {journal} {Phys. Rev. C}\ }\textbf {\bibinfo {volume} {88}},\
  \bibinfo {pages} {044915} (\bibinfo {year} {2013})},\ \Eprint
  {http://arxiv.org/abs/1301.4470} {arXiv:1301.4470 [hep-ph]} \BibitemShut
  {NoStop}%
\bibitem [{\citenamefont {Lisa}\ \emph
  {et~al.}(2000{\natexlab{c}})\citenamefont {Lisa}, \citenamefont {Ajitanand},
  \citenamefont {Alexander} \emph {et~al.}}]{LISA20001}%
  \BibitemOpen
  \bibfield  {author} {\bibinfo {author} {\bibfnamefont {M.}~\bibnamefont
  {Lisa}}, \bibinfo {author} {\bibfnamefont {N.}~\bibnamefont {Ajitanand}},
  \bibinfo {author} {\bibfnamefont {J.}~\bibnamefont {Alexander}},  \emph
  {et~al.},\ }\href {\doibase https://doi.org/10.1016/S0370-2693(00)01280-6}
  {\bibfield  {journal} {\bibinfo  {journal} {Physics Letters B}\ }\textbf
  {\bibinfo {volume} {496}},\ \bibinfo {pages} {1} (\bibinfo {year}
  {2000}{\natexlab{c}})}\BibitemShut {NoStop}%
\end{thebibliography}%

      \end{document}